\documentclass[10pt,twocolumn]{article}
\usepackage{hyperref} 
\usepackage{authblk}

\usepackage{geometry}
\usepackage{graphicx}
\usepackage{mathrsfs}
\usepackage{mathtools, cuted}
\usepackage{float}

\title{\textbf{Submegahertz spectral width photon pair source based on fused silica microspheres}}

\author[1]{Erasto Ortiz-Ricardo}
\author[1]{Cesar Bertoni-Ocampo}
\author[1]{Mónica Maldonado-Terrón}
\author[1]{Arturo Sanchez Zurita}
\author[2]{Roberto Ramirez-Alarcon}
\author[1]{Hector Cruz Ramirez}
\author[3]{R. Castro-Beltrán}
\author[1,*]{Alfred B. U'Ren}

\affil[1]{\small Instituto de Ciencias Nucleares, Universidad Nacional Aut\'onoma de M\'exico, Apartado Postal 70-543, 04510 DF, M\'exico}
\affil[2]{Centro de Investigaciones en \'Optica A.C., Loma del Bosque 115, Colonia Lomas del Campestre, 37150 Le\'on Guanajuato, M\'exico}
\affil[3]{Departamento de Ingeniería Física, Cuerpo Académico de Mécanica Estadística, Divisón de Ciencias e Ingenierías, Universidad de Guanajuato, León, Guanajuato, 37150, Mexico}
\affil[*]{Corresponding author: alfred.uren@correo.nucleares.unam.mx}

\date{\small October, 2021}

\PassOptionsToPackage{%
	backend=bibtex8,bibencoding=utf8,%
	language=auto,%
	style=numeric-comp,%
	sorting=none, 
	maxbibnames=10, 
	natbib=true 
}{biblatex}
\usepackage{biblatex}

\begin{document}
\twocolumn[
\begin{@twocolumnfalse}
	\maketitle
	\begin{abstract}
	High efficiency, sub-MHz bandwidth photon pair generators will enable the field of quantum technology to transition from laboratory demonstrations to transformational applications involving information transfer from photons to atoms. While spontaneous parametric processes are able to achieve high efficiency photon pair generation, the spectral bandwidth tends to be relatively large, as defined by  phasematching constraints. To solve this fundamental limitation, we use an ultra-high quality factor ($Q$) fused silica microsphere resonant cavity to form a photon pair generator.  We present the full theory
	for the SFWM process in these devices, fully taking into account all relevant source characteristics in our experiments.   The exceptionally narrow (down to kHz-scale) linewidths of these devices in combination with the device size results in a reduction in the bandwidth of the photon pair generation, allowing sub-MHz spectral bandwidth to be achieved. Specifically, using a pump source centered around 1550nm, photon pairs with the signal and idler modes at wavelengths close to 1540nm and 1560nm, respectively, are demonstrated. We herald a single idler-mode photon by detecting the corresponding signal photon, filtered via transmission through a wavelength division multiplexing channel of choice.  We demonstrate the extraction of the spectral profile of a single peak in the single-photon frequency comb from a measurement of the signal-idler time of emission distribution. These improvements in device design and experimental methods enabled the narrowest spectral width ($\Delta\nu=366$kHz) to date in a heralded single photon source based on SFWM. 
	\end{abstract}
\end{@twocolumnfalse}
]

\section{Introduction}
\label{sec:introduction}
Advances in quantum technologies over the past two decades have made possible an exciting breadth of applications in fields such as communications~\cite{gisin2007quantum}, imaging~\cite{moreau2019imaging}, and computation~\cite{ladd2010quantum}. Photon pair generation based on the spontaneous parametric downconversion~\cite{burnham1970observation} and four wave mixing~\cite{sharping2001observation}  processes have played an essential role in this revolution due to the ease with which the quantum entanglement characteristics of the emitted signal and idler photons may be tailored, and on account of their ability to propagate long distances either in free space or in optical fibers with minimal interaction with the environment. However, a number of key challenges must be overcome in order for photon pair generation technology to achieve its true potential, including: i) source miniaturization enabling the eventual on-chip integration of source, optical manipulation, and detection \cite{Moody_2020}, \cite{Gaeta2020} ii) increasing the conversion efficiency, thus permitting high-brightness photon-pair emission with the lowest possible pump power, iii) photon pair indistinguishability, including spectral factorizability, permitting photons from distinct sources to interfere~\cite{uren2005generation}, and iv) the reduction of the emission bandwidth to the MHz, or sub-MHz, level so as to be compatible with atomic electronic transitions\cite{Liu_Jianji}. The last requirement must be achieved in order to create single atom-single photon interfaces which will facilitate information transfer from photons, in the form of flying qubits, to atoms thus constituting a quantum memory. This technology is essential for the further progress of quantum information processing based on photons.

An optical platform which has the potential to simultaneously meet all of these requirements is the ultra-high-$Q$ optical microresonator. As a result of the long photon lifetimes inside the cavities, extremely high circulating intensities are possible, resulting in very low optical thresholds for nonlinear behaviors \cite{Vahala2003microcavities,ji2017ultra,shen2018low}.  In addition, the extremely narrowband resonances lead to nonlinear optical effects occurring at very well-defined frequencies.  Because cavity-enhanced photon pair generation involves emission in well-defined cavity modes, by isolating a single cavity resonance for each of the signal and idler modes, the resulting two-photon state ends up being naturally spectrally factorizable, ensuring indistinguishability\cite{garay2012theory}. Lastly, microresonators are uniquely well-positioned to permit source integration\cite{reimer2015cross, kues2017chip}.

There are two possible routes for generating photon-pairs based on spontaneous parametric processes which can occur in microresonators:  the spontaneous parametric down conversion (SPDC) process based on second-order non linear materials  \cite{pomarico2009waveguide, guo2017parametric, furst2011quantum, fortsch2013versatile, hockel2011direct, luo2017temporal}, and  the spontaneous four wave mixing (SFWM) process based on third order nonlinear materials \cite{grassani2015micrometer, rogers2016high,jaramillo2017persistent,preble2015chip,lu2016biphoton,lu2019chip,silverstone2015qubit,caspani2016multifrequency}. Photon pair generation from a cavity-enhanced source,  in which the non-linear medium is contained by an optical cavity, based on either of these two processes with a narrowband pump leads to a joint spectral amplitude in the form of a frequency comb expressed as a function of the frequency difference  $\omega_s-\omega_i$, in terms of the signal $\omega_s$ and idler $\omega_i$ frequencies.    While each resulting comb peak has a width which is inversely proportional to the cavity quality factor $Q$ , the intra-peak separation (or free spectral range) is inversely proportional to the cavity round-trip time\cite{garay2012theory,moreno2010theory}.   Thus, an important benefit of a microresonator as compared to an extended cavity design is that the comb peaks end up being separated by a greater spectral distance facilitating the possibility of addressing individual comb peaks.   Note that in the classical nonlinear optics realm, four wave mixing in optical resonators is likewise known to naturally give rise to the emission of frequency combs, with applications in the field of atomic clocks and generally in metrology\cite{scott2014clock,drake2019terahertz}.

Note that in the standard  SPDC and SFWM processes (without the use of optical cavities), the spread of emission frequencies is limited mainly by phase matching constraints and can be substantial.   Such large bandwidths are useful in certain situations, e.g. they lead to narrow Hong-Ou-Mandel interference dips which in turn permit a large resolution in quantum optical coherence tomography devices\cite{Teich2012tomography,graciano2019tomography}.   However, as already mentioned, the basic requirement for the development of single atom - single photon interfaces, is the emission of narrowband photon pairs\cite{wang2019memory,schunk2015interfacing,wolfgramm2011atom}.  

Cavity-enhanced SPDC sources have been demonstrated using integrated microresonators \cite{guo2017parametric,fortsch2013versatile},  nonlinear waveguide cavities \cite{luo2017temporal}, and free-space extended cavities\cite{rambach2016sub,hockel2011direct,ou1999cavity,slattery2015narrow,fekete2013ultranarrow}. In the case of SFWM, cavity-enhanced photon-pair sources have likewise been based on microring\cite{reimer2015cross,grassani2016energy,kues2017chip,jaramillo2017persistent,preble2015chip,lu2019chip,silverstone2015qubit,caspani2016multifrequency} and microdisk\cite{rogers2019coherent} cavities.  In this paper we report the first demonstration of a photon-pair source based on fused silica microspheres, and  present a theory for SFWM in these devices which leads to simulations which agree well with our measurements.  In this work we extend our previous cavity-enhanced SFWM theory\cite{garay2012theory}, so as to include important characteristics relevant to our current experimental results such as: i) all four waves at resonance in the cavity, ii) pump varied in time to maintain resonance, leading to two-photon state in the form of a statistical mixture, and iii) analysis carried out for micro- rather than extended cavities. Here we report ultra-narrow photon-pair generation, with emission bandwidths down to $\Delta \nu $=366kHz. While this small single-photon bandwidth is similar to those observed in certain extended-cavity SPDC sources (e.g. 666kHz in \cite{rambach2016sub} and 265kHz in  \cite{Liu_Jianji}), it represents a $\sim \times 43$ improvement with respect the previously-reported spectrally narrowest SFWM source (15.9MHz)\cite{imany201850}.

\section{Theory for the SFWM process in microspheres}
\label{sec:theory}

\begin{figure}[ht]
	\centering
	\includegraphics[width=8cm]{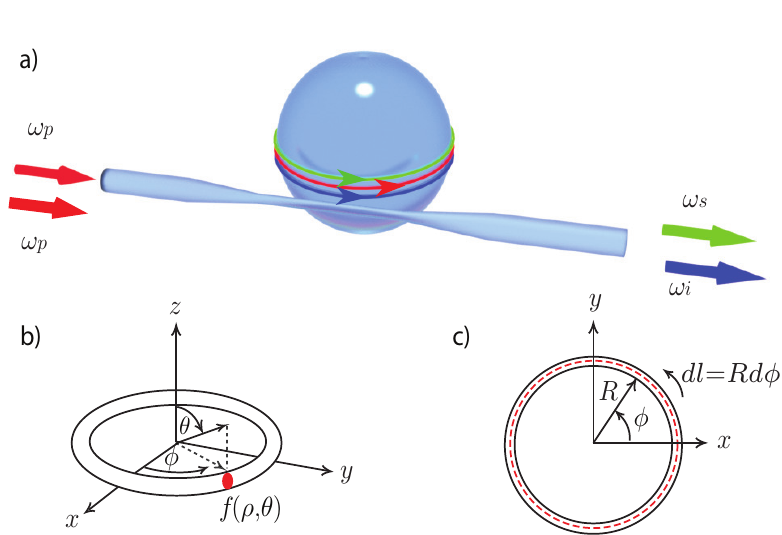}
	\caption{\label{fig:Fourier} a) Schematic of a SFWM photon-pair source based on a fused silica microsphere, evanescently coupled to a fiber taper placed in closed proximity.
		b) Transverse mode of propagation around of the sphere perimeter. c) Top view of the guided mode. }
\end{figure}

Here we are interested in studying photon pairs produced by SFWM in a fused silica microsphere, with radius $R$, as the nonlinear medium.  The pump is assumed to be coupled evanescently from an elongated (tapered) fiber
to a mode circulating on the sphere's equator; see for example Ref. \cite{Cai2000critical}.  Photon pairs produced in the sphere can then couple out back to the tapered fiber, whence they may be directed as desired to an experiment of interest; see Fig. \ref{fig:Fourier}.   

Each of the four waves participating in the SFWM process is assumed to propagate along the equator on the plane $\theta=\pi/2$,   parallel to the unit vector  $\vec{e}_{\phi}$.  Expressed in spherical coordinates, the Hamiltonian for this process may be written as

\begin{align}
	\label{Hamiltonian}
	\hat{H}(t)& =\frac{3}{4}\epsilon_0\chi^{(3)}\mkern-2.5mu  \!\!\int\limits_0^{2 \pi}\mkern-7mu  d\phi \!\!\int\limits_0^{ \pi} \mkern-7mu d\theta  \!\!\int\limits_0^R \mkern-7mu d\rho  \rho^{2} \sin \theta  E_{1}^{(+)} (\vec{r},t)\ E_{2}^{(+)} (\vec{r},t) \\ \nonumber
	&\times \hat{E}_{s}^{(-)} (\vec{r},t) \hat{E}_{i}^{(-)} (\vec{r},t)+H.C..
\end{align}

\noindent in terms of the third-order optical nonlinearity $\chi^{(3)}$, the permittivity of free-space $\epsilon_0$, the positive-frequency electric field operators for each of the two pumps $E_{1}^{(+)}(\vec{r},t)$ and $E_{2}^{(+)}(\vec{r},t)$, and the negative-frequency electric field operators for the signal and the idler, $\hat{E}_{s}^{(-)}(\vec{r},t)$ and $\hat{E}_{i}^{(-)}(\vec{r},t)$.

We describe each mode supported by the sphere by an index vector $\vec{l}$ composed of the $l$, $m$, and $q$ individual indices related to azimuthal, polar and radial coordinates respectively.  Each of the pump waves  ($\nu=1$ and $\nu=2$), traveling in modes $\vec{l}_1$ and $\vec{l}_2$, respectively, is described classically  as 

\begin{equation}
	\label{clasical_field}
	E_{\nu}^{(+)}(\vec{r},t)=\sum_{\vec{l}_\nu} A_{\vec{l}_\nu}f_{\vec{l}_\nu}(\rho,\theta)\int d\omega \alpha_{\vec{l}_\nu}  \ (\omega)e^{-i\left[\omega t-k_{\vec{l}_\nu}(\omega )R \phi\right]}
\end{equation}

\noindent where $A_{\vec{l}_\nu}$ is the amplitude, $f_{\vec{l}_\nu} \ (\rho,\theta)$ the transverse field distribution  (for a fixed value of $\phi$), and $\alpha_{\vec{l}} \ (\omega)$ is the spectral envelope for each of the two pump waves, $\nu=1$ and $\nu=2$.   The electric field for the signal and idler photons are described quantum mechanically as follows

\begin{equation}
	\label{quantum_field}
	\hat{E}_{\nu}^{(+)}(\vec{r},t)=\sum_{\vec{l}_\nu} f_{\vec{l}_\nu}(\rho,\theta) \int d\omega \ell(\omega)e^{-i\left[\omega t-k_{\vec{l}_\nu}(\omega)R \phi\right]}\hat{b}_{\vec{l}_\nu} (k_{\omega})
\end{equation}

\noindent where $\hat{b}_{\vec{l}_\nu} (k_{\omega})$ is the annihilation operator for mode $\vec{l}_\nu$, for each of the signal ($\nu=s$)  and  idler ($\nu=i$);  $f_{\vec{l}_\nu} \ (\rho,\theta)$ represents the signal and idler transverse spatial distributions,  and the function $ \ell(\omega)$ is given as $\ell(\omega)=\omega^{1/2}/[n(\omega)v_{g}(\omega)]$, with $n(\omega)$ and $v_{g}(\omega)$ the refractive index and the group velocity, respectively.

The resonant wavelengths  in the sphere  can be obtained solving numerically the following equation, derived from a Mie scattering approach, for $\lambda$ (for particular values of the indices azimuthal $l$, polar $m$ and radial $q$). \cite{Armani_dispersion}  

\begin{equation} \label{E:disp}
	\begin{split}
		\frac{1}{\lambda } &  =
		\frac{1}{2\pi R n_{s}(\lambda)}\left( \nu+2^{-\frac{1}{3}} \alpha_{q}\nu^{\frac{1}{3}}-\frac{P(\lambda)}{(n_{s}(\lambda)^{2}-1)^\frac{1}{2}} \right.   + \\
		& \frac{3}{10}2^{-\frac{2}{3}}\alpha_{q}^{2}\nu^{-\frac{1}{3}} 
		-\frac{2^{-\frac{1}{3}}P(\lambda)(n_{s}(\lambda)^{2}-\frac{2}{3}P(\lambda)^{2})}{(n_{s}^{2}(\lambda)-1)^{\frac{3}{2}}}\alpha_{q}\nu^{-\frac{2}{3}} \bigg),
	\end{split}
\end{equation}

\noindent where $R$ is the sphere radius, $P(\lambda)=n_s(\lambda)$, for a TE mode, and $P(\lambda)=1/n_s(\lambda)$, for a TM mode,  with $n_{s}(\lambda)$ the fused silica index of refraction obtained from the Sellmeier equation \cite{agrawal2006pulse}. In \eqref{E:disp},  $\nu=l+1/2$ and $\alpha_{q}$ represents the zeros of the Airy function $\mbox{Ai}(-z)$; for simplicity, in this work we have considered only modes with  $l=m$ and $q=1$, so that $\vec{l}=(l,m,q)$ reduces to $(l,l,1)$.  We can obtain the effective index for each WG mode $n_{eff}(\lambda)$, with help of the resonance condition,   as $n_{eff}(\lambda)=l \lambda/L$, with $L=2\pi R$. 
The dispersion relation is subsequently obtained as  $k(\omega)=n_{eff}(\omega)\omega/c$.

The quantum state is then obtained, following a standard perturbative approach, as

\begin{equation}
	\vert \Psi \rangle \approx \vert 0 \rangle + 
	\frac{1}{i \hbar} 
	\int\limits_0^t dt' \hat{H}(t') \vert 0 \rangle 
\end{equation}

Replacing the expressions for the electric field (equations (\ref{clasical_field}) and (\ref{quantum_field})) into the Hamiltonian (equation \ref{Hamiltonian}), we obtain the quantum state $\vert \Psi \rangle = \vert 0 \rangle + \eta \vert \Psi_2 \rangle$, written in terms of a constant related to the source brightness $\eta$ and the two-photon component of the state   $\vert \Psi_2 \rangle$, the latter given by

\begin{equation}
	\begin{split}
		\vert \Psi_2 \rangle=&\frac{1}{i \hbar} \int\limits_0^t d t' \hat{H}(t') \vert 0 \rangle  \\
		=&  \sum_{\vec{l}_1,\,\vec{l}_2,\,\vec{l}_s,\,\vec{l}_i} A_{\vec{l}_1} A_{\vec{l}_2} \Theta_{\vec{l}_1 \vec{l}_2 \vec{l}_s \vec{l}_i}  \int d\omega_{s} \int d\omega_{i}  \biggl[ l(\omega_{s})l(\omega_{i})   \\ 
		\times & \int d\omega_{1} \biggl(\alpha_{\vec{l}_1} (\omega_{1})\alpha_{\vec{l}_2} (\omega_{s}+\omega_{i}-\omega_{1}) 
		G_{\vec{l}_1 \vec{l}_2 \vec{l}_s \vec{l}_i}(\omega_1,\omega_s,\omega_i)\\
		&\times \hat{b}_{\vec{l}_s} ^{\dagger}(\omega_{s})\hat{b}_{\vec{l}_i}^{\dagger}(\omega_{i}) \vert 0 \rangle \biggl)\biggl].    
	\end{split}
\end{equation}


Here we have assumed that the time interval between photon-pair emission events is much greater than the characteristic time for each event, so that the limits of the temporal integral may be extended to $\pm \infty$.   Note that in writing this expression for $\vert \Psi_2 \rangle$, we have defined the field overlap $ \Theta_{\vec{l}_1 \vec{l}_2 \vec{l}_s \vec{l}_i}$, given by 

\begin{equation}
	\label{E:traslape}
	\Theta_{\vec{l}_1 \vec{l}_2 \vec{l}_s \vec{l}_i}=\int d\rho\int d\theta \rho^{2}\sin\theta f_{1}(\rho,\theta)f_{2}(\rho,\theta)f_{s}^{*}(\rho,\theta)f_{i}^{*}(\rho,\theta),
\end{equation}

\noindent and the function  $G_{\vec{l}_1 \vec{l}_2 \vec{l}_s \vec{l}_i}(\omega_1,\omega_s,\omega_i)$  expressed in terms of the sphere's equatorial perimeter $L= 2 \pi R$, as

\begin{multline}
	\label{E:intramodalPM}
	G_{\vec{l}_1 \vec{l}_2 \vec{l}_s \vec{l}_i}(\omega_1,\omega_s,\omega_i)=\\ \mbox{sinc}\left(\frac{L  \Delta k_{\vec{l}_1 \vec{l}_2 \vec{l}_s \vec{l}_i}(\omega_1,\omega_s,\omega_i)}{2}\right)\exp\left(i\frac{L  \Delta k_{\vec{l}_1 \vec{l}_2 \vec{l}_s \vec{l}_i}(\omega_1,\omega_s,\omega_i)}{2} \right),
\end{multline}

\noindent defined in turn in terms of the phasemismatch function 

\begin{equation}
	\Delta k_{\vec{l}_1 \vec{l}_2 \vec{l}_s \vec{l}_i}(\omega_1,\omega_s,\omega_i)=k_{\vec{l}_1}(\omega_1)\!+\!k_{\vec{l}_2}(\omega_s\!+\!\omega_i\!-\!\omega_1)\!-\!k_{\vec{l}_s}(\omega_s)\!-\!k_{\vec{l}_i}(\omega_i).
\end{equation}

Let us now assume that the two pumps are \emph{degenerate} (spatially and spectrally), as well as  \emph{monochromatic} at frequency $\omega_p$, i.e.

\begin{equation}
	\alpha_{\vec{l}_p}(\omega)\equiv \alpha_{\vec{l}_1}(\omega) =\alpha_{\vec{l}_2} (\omega)=\delta (\omega-\omega_p).
\end{equation}

Let us consider a short longitudinal section of the continuous-wave pump of length $\Delta z_p$ (with $\Delta z_p \ll L$).   Performing the change of variables $\Omega=\omega_s-\omega_p$ we then obtain the state resulting from one cavity round trip of this longitudinal pump section of $\Delta z_p$ length 

\begin{align}
	\label{E:2Pstate}
	\vert \Psi_2 \rangle  &=  \sum_{\vec{l}_p,\,\vec{l}_s,\,\vec{l}_i} A^2_{\vec{l}_p} \Theta_{\vec{l}_p \vec{l}_p \vec{l}_s \vec{l}_i}  \int d\Omega  \ell(\omega_{p}+\Omega)\ell(\omega_{p}-\Omega) \times  \\ \nonumber
	&\times 
	g_{\vec{l}_p \vec{l}_s \vec{l}_i}(\Omega)
	\hat{b}_{\vec{l}_s} ^{\dagger}(\omega_{p}+\Omega)\hat{b}_{\vec{l}_s} ^{\dagger}(\omega_p-\Omega) \vert 0 \rangle,
\end{align}

\noindent where the function $g_{\vec{l}_p \vec{l}_s \vec{l}_i}(\Omega)$, which constitutes a reduced version of $G_{\vec{l}_1 \vec{l}_2 \vec{l}_s \vec{l}_i}(\omega_1,\omega_s,\omega_i)$,  can be expressed as

\begin{multline}
	g_{\vec{l}_p \vec{l}_s \vec{l}_i}(\omega_{p},\Omega)=\\
	\mbox{sinc}\left(\frac{L  \Delta \kappa_{\vec{l}_p  \vec{l}_s \vec{l}_i}(\omega_{p},\Omega)}{2}\right)\exp\left(i\frac{L  \Delta \kappa_{\vec{l}_p \vec{l}_s \vec{l}_i}(\omega_{p},\Omega)}{2} \right)
\end{multline}

\noindent in terms of the reduced phasemismatch function 

\begin{equation}
	\Delta \kappa_{\vec{l}_p  \vec{l}_s \vec{l}_i}(\omega_{p},\Omega)= 2 k_{\vec{l}_p}(\omega_p)\!-\!k_{\vec{l}_s}(\omega_p\!+\!\Omega)\!-\!k_{\vec{l}_i}(\omega_p\!-\!\Omega)\!-\!2\gamma P.
\end{equation}

Note that in the above expression we have incorporated a nonlinear term in the phasemismatch associated with self and cross phase modulation $2 \gamma P$ \cite{agrawal2006pulse}.  This term can be written in terms of the $Q$ parameter of the cavity, of the input power $P_{in}$, of the nonlinear index of refraction $n_2$, and of the effective transverse mode area $A_{eff}$ as

\begin{equation}
	\label{equ:gamma}
	2 \gamma P = \dfrac{P_{in} Q n_{2}}{\pi nR A_{eff}}.
\end{equation}

Using the fact that the function $\ell(\omega)$ is a slow function of $\omega$, and assuming that each of the three waves (the degenerate pump, signal, and idler) each travel in a single spatial mode, the quantum state can be simplified as $\vert \Psi \rangle= \vert 0 \rangle + \eta' \vert \Psi_2 \rangle$ where the new constant $\eta'$ incorporates the quantity $A^2_{\vec{l}_p} \Theta_{\vec{l}_p \vec{l}_p \vec{l}_s \vec{l}_i}$, with  $\vert \Psi_2 \rangle$  given by

\begin{equation}
	\vert \Psi_2 \rangle  =
	\int d\Omega  
	g_{\vec{l}_p \vec{l}_s \vec{l}_i}(\Omega)
	\hat{b}_{\vec{l}_s} ^{\dagger}(\omega_{p}+\Omega)\hat{b}_{\vec{l}_s} ^{\dagger}(\omega_p-\Omega)
	\vert \mbox{vac} \rangle.
\end{equation}
So far, the quantum state has been expressed in terms of the annihilation operators $\hat{b}_{\vec{l}_s}$ and $\hat{b}_{\vec{l}_i}$ which correspond to the modes resonating within the sphere.    In any realistic experiment we of course need  to couple the photon pairs out of the resonating sphere so as to use them in an experimental setup of interest.   This can be accomplished through evanescent coupling of the sphere modes to a given propagation mode of a tapered fiber placed in close proximity to the sphere.  Let us denote as 
$\hat{a}_\nu$   
the extra-cavity mode in the tapered fiber (we assume that light couples into a single  taper mode)  for the signal ($\nu=s$) and idler ($\nu=i$), with $r_\nu$ representing the reflectivity of the sphere-taper interface (the probability amplitude corresponding to a photon remaining within the sphere), and $t'_\nu$ representing the transmissivity (the probability amplitude corresponding to a photon coupling from the sphere to the taper); in a  lossless cavity, energy conservation dictates $\vert t'_\nu\vert ^{2}+\vert r_\nu\vert ^{2}=1$, with $\vert r_{\nu}\vert =1-\frac{l\pi}{Q}$. After $n+1$ iterations in the cavity the intra-cavity mode $\hat{b}_{\vec{l}_s}$ and $\hat{b}_{\vec{l}_i}$ can be expressed as follows

\begin{multline}
	\hat{b}_{\vec{l}_\nu} ^{\dagger}(\omega) \rightarrow
	r_\nu^{n+1}e^{i n k_{\vec{l}_\nu}(\omega)L}  \hat{a}_\nu^{\dagger}(\omega)\\+
	\left(
	\frac{1-r_\nu^{n+1}e^{i(n+1) k_{\vec{l}_\nu (\omega)}  L}}
	{1-r_\nu e^{i k_{\vec{l}_\nu} (\omega)  L}}
	\right)
	t'_\nu
	\hat{a}_\nu^{\dagger}(\omega)
\end{multline}

Taking the limit $n \rightarrow \infty$, so that all SFWM light produced by a single  cavity round trip of the longitudinal pump section of length $\Delta z_p$ is allowed to escape the cavity,  we may write the extra-cavity mode operator as follows

\begin{equation}
	\lim_{n \to \infty} \hat{b}_{\vec{l}_\nu} ^{\dagger}(\omega)\vert 0 \rangle_\nu =A_\nu(\omega) \hat{a}_\nu^{\dagger}(\omega) \vert 0 \rangle_\nu 
\end{equation}

\noindent written in terms of the Airy function $A_\nu(\omega)$

\begin{equation}
	A_\nu(\omega) =\frac{t'_\nu}
	{1-r_\nu e^{i k_{\vec{l}_\nu}(\omega)  L}}.
\end{equation}

We may then write the two-photon state  propagating in the tapered fiber modes $\vert \Psi_2'\rangle$, described by annihilation operators $ \hat{a}_s$ and $ \hat{a}_i$, as 

\begin{equation}
	\begin{split}
		\vert \Psi_2' \rangle  &=
		\int d\Omega  
		g_{\vec{l}_p \vec{l}_s \vec{l}_i}(\Omega)
		A_s(\omega_p+\Omega) A_i(\omega_p-\Omega)
		\times \\
		\times &
		\hat{a}_s ^{\dagger}(\omega_{p}+\Omega)\hat{a}_i ^{\dagger}(\omega_p-\Omega)
		\vert 0 \rangle.
	\end{split}
\end{equation}

Let us note that this is the quantum state produced by a single pass of the pump field through the cavity.   In an experimental situation of interest the pump field is resonant in the cavity, so that with a probability  amplitude  $t$ each pump photon couples from the taper to the sphere, and once within the cavity it remains  with a probability amplitude  $r_p$ at each pass through the taper-sphere interface.    We can then write an expression for the two-photon state, resulting from $n$ iterations of the pump in the cavity $\vert \Psi_2 ''\rangle$

\begin{equation}
	\begin{split}
		\vert \Psi_2'' \rangle &=
		t_p \vert \Psi_2'\rangle + 
		t_p r_p^2 e^{i  k_{\vec{l}_p}(\omega_p) L }  \vert \Psi_2' \rangle \\ 
		&+t_p r_p^4 e^{i  2 k_{\vec{l}_p}(\omega_p)  L } \vert \Psi_2' \rangle + \cdots  \\
		&\cdots+
		t_p r_p^{ 2n} e^{i 2 n k_{\vec{l}_p}(\omega_p)  L } \vert \Psi_2'\rangle \\
		&=
		\frac{
			t_p[1-r_p^{2(n+1)}e^{i  k_{\vec{l}_p}(\omega_p) L(n+1) }]
		}
		{
			1-r_p^2 e^{i  k_{\vec{l}_p}(\omega_p) L}
		}
		\vert \Psi_2'\rangle
	\end{split}
\end{equation}
In this expression, upon each successive round trip of the pump longitudinal section $\Delta z_p$  in the cavity, the  electric-field amplitudes $A_{\vec{l}_1}$ and $A_{\vec{l}_2}$ are each reduced by $r_p$, so that a factor $r_p^2$ appears in addition to the phase term corresponding to one round trip for each pump  (both from a single degenerate pump mode).  In the limit $n \rightarrow \infty$ corresponding to a situation in which all the pump light from longitudinal section $\Delta z_p$ initially coupled into the cavity has escaped, we may then write the resulting two-photon state $\vert\Psi_2''\rangle$ as follows

\begin{equation}
	\lim_{n \to \infty} \vert \Psi'' \rangle = A_p ( \omega_p) \vert \Psi' \rangle 
\end{equation}

\noindent in terms of the Airy function for the pump $A_p(\omega)$ expressed as

\begin{equation}
	A_p(\omega) =\frac{t_p}
	{1-r_p^2 e^{i  k_{\vec{l}_p}(\omega)  L}}
	\label{eq:Airy_pump}
\end{equation}

We can then write the two photon-state, which  incorporates the full effect of the cavity,  as follows\footnote{Note that the derivation shown here pertains to the case of degenerate pump waves.  In the case of non-degenerate pumps (spatially and/or spectrally), two separate Airy functions will appear, one for each of the pumps.}
\begin{equation}
	\begin{split}
		\vert \Psi''(\omega_p) \rangle & =  \vert \mbox{vac} \rangle +  \eta'   
		A_p(\omega_p) \times \\ & \times
		\int d\Omega  
		f(\Omega;\omega_p)
		\hat{a}_s ^{\dagger}(\omega_{p}+\Omega)
		\hat{a}_i ^{\dagger}(\omega_p-\Omega)
		\vert \mbox{vac} \rangle
		\\ 
		& \equiv  \vert \mbox{vac} \rangle +  \eta'   
		A_p(\omega_p)
		\vert \varphi(\omega_p) \rangle,
	\end{split}
\end{equation}

\noindent in terms of the joint spectral amplitude function $f(\Omega;\omega_p)$

\begin{equation}
	f(\Omega; \omega_p)=g_{\vec{l}_p \vec{l}_s \vec{l}_i}(\Omega; \omega_p)
	A_s(\omega_p+\Omega) A_i(\omega_p-\Omega).
\end{equation}

Alternatively, it is useful for visualization purposes to write the two photon state in an equivalent form involving the two-dimensional frequency generation space $\{\omega_s,\omega_i\}$, as follows

\begin{multline}
	\label{Eq:2dJSI}
	\vert \Psi''(\omega_p) \rangle  =\vert \mbox{vac} \rangle +\\+ \eta'  A_p(\omega_p)
	\int d\omega_s 
	\int d\omega_i  
	f_2(\omega_s,\omega_i;\omega_p)
	\hat{a}_s ^{\dagger}(\omega_s)
	\hat{a}_i  ^{\dagger}(\omega_i)
	\vert \mbox{vac} \rangle.
\end{multline}

\noindent in terms of the two-dimensional joint spectral amplitude

\begin{multline}
	f_2(\omega_s,\omega_i;\omega_p)=\\=
	\delta(\omega_{s}+\omega_{i}-2 \omega_{p})  
	G_{\vec{l}_p \vec{l}_p \vec{l}_s \vec{l}_i}(\omega_p,\omega_s,\omega_i)
	A_s(\omega_s)
	A_i(\omega_i).
\end{multline}

It is convenient to re-express this joint amplitude in terms of `rotated' variables: $\Omega=(\omega_s-\omega_i)/2$ (already introduced) and $\omega_{+}=(\omega_s+\omega_i)/2$ as

\begin{multline}
	f'_2(\omega_+,\Omega;\omega_p)=
	\delta(\omega_{+}- \omega_{p})\times \\ \times
	G_{\vec{l}_p \vec{l}_p \vec{l}_s \vec{l}_i}(\omega_p,\omega_++\Omega,\omega_+-\Omega)
	A_s(\omega_++\Omega)
	A_i(\omega_+-\Omega).
\end{multline}

In our experiments, while the pump wave is in the form of a continuous wave with a narrowband linewidth $\delta_p$, the pump frequency is varied in time according to a triangular wave with amplitude $\Delta_p$ and frequency $f_p$.      As $\omega_p$ is swept within the spectral window of width $\Delta_p$, a bandwidth $\Delta_r$ corresponding to the spectral width of the cavity resonance function $\vert A_p(\omega_p)\vert^2$ can couple into the cavity. There are therefore three bandwidths of interest which govern the pump and its interaction with the cavity: i) pump linewidth $\delta_p$, ii) pump resonance bandwidth $\Delta_p$, and iii) pump sweeping range $\Delta_r$.  In our experiments (see below) the relationships $\delta_p \ll \Delta_p \ll \Delta_r $ are fulfilled with approximate values $\delta_p \approx 200$kHz, $\Delta_p \approx$ 20 MHz, and $\Delta_r \approx 25$GHz.

Let us now express the two-photon state produced by the sphere as a statistical mixture of all the pure states produced by each individual pump spectral component $\omega_p$ which can couple into the sphere, as follows

\begin{equation}
	\hat{\rho}=
	\int d \omega_p \vert A_p(\omega_p) \vert^2
	\vert \varphi(\omega_p) \rangle
	\langle \varphi(\omega_p) \vert
\end{equation}

We can now write down an expression for the the spectral intensity (SI) for the idler photon $R_i(\Omega)$, in terms of the photon number operators $\hat{n}(\omega_\nu)=a^\dagger(\omega_\nu) a(\omega_\nu)$ (with $\nu=s,i$),  as follows

\begin{align}
	R_i(\Omega)&=
	\int d\omega_s
	\langle \hat{n}(\omega_s) \hat{n}(\omega_p-\Omega) \rangle
	\\ 
	& = \mbox{Tr}\left( 
	\hat{a}^\dagger(\omega_s)
	\hat{a}^\dagger(\omega_p-\Omega)
	\hat{a}(\omega_p-\Omega)
	\hat{a}(\omega_s)
	\hat{\rho} 
	\right) \nonumber
	\\ 
	&=\int d \omega_p \vert A(\omega_p) f(\Omega ; \omega_p ) \vert^2 
\end{align}

An explicit version of the above equation is as follows, where in the second line we have used the approximation that the state is defined by the three Airy functions with a negligible effect of the phasematching function.

\begin{align}
	\label{JSIfinal}
	R_i(\Omega)&=
	\int d \omega_p \vert 
	g_{\vec{l}_p \vec{l}_s \vec{l}_i}(\Omega) \vert^2
	\mathscr{A}_p(\omega_p)
	\mathscr{A}_s(\omega_p+\Omega)
	\mathscr{A}_i(\omega_p-\Omega)\nonumber
	\\ 
	&\approx
	\int d \omega_p 
	\mathscr{A}_p(\omega_p)
	\mathscr{A}_s(\omega_p+\Omega)
	\mathscr{A}_i(\omega_p-\Omega)
\end{align}

\noindent in terms of the intensity Airy functions $\mathscr{A}_\nu(\omega)\equiv\vert A_\nu(\omega) \vert^2$ (with $\nu=p,s,i)$.
It is of interest to write equivalent expressions in time domain, for which we define  time-domain annihilation operators as follows

\begin{equation}
	\hat{a}_t(t)=\frac{1}{2 \pi} \int d \omega \hat{a}(\omega) e^{-i \omega t}.
\end{equation}  

We can then  write down an expression for the resulting two-photon times of emission distribution (TED) $\tilde{R}(t_s,t_i)$ as

\begin{align}
	\label{E:TED}
	\tilde{R} (t_s,t_i)&=
	\langle \hat{n}_t(t_s) \hat{n}_t(t_i) \rangle=
	\mbox{Tr}\left( 
	\hat{a}_t^\dagger(t_s)
	\hat{a}_t^\dagger(t_i)
	\hat{a}_t(t_i)
	\hat{a}_t(t_s)
	\hat{\rho} 
	\right) \nonumber
	\\ 
	&=
	\int d\omega_p
	\mathscr{A}_p(\omega_p) 
	\int d \Omega 
	\vert f(\Omega ; \omega_p ) \vert^2
	e^{-i(t_s-t_i) \Omega} \nonumber
	\\ 
	&=
	\int d \Omega
	\int d\omega_p
	\mathscr{A}_p(\omega_p)
	\vert f(\Omega ; \omega_p ) \vert^2
	e^{-i(t_s-t_i) \Omega} \nonumber
	\\ 
	&=
	\int d \Omega
	R_i(\Omega)
	e^{-i(t_s-t_i) \Omega}, 
\end{align}

Note that in the third line, we have interchanged the order of integration, leading to a Fourier transform relationship between the SI and the TED (fourth line)

\section{Specific example: illustration of the spectral / temporal photon-pair properties}
\label{sec:example}

\begin{figure*}[!ht]
	\centering
	\includegraphics[width=13.6cm]{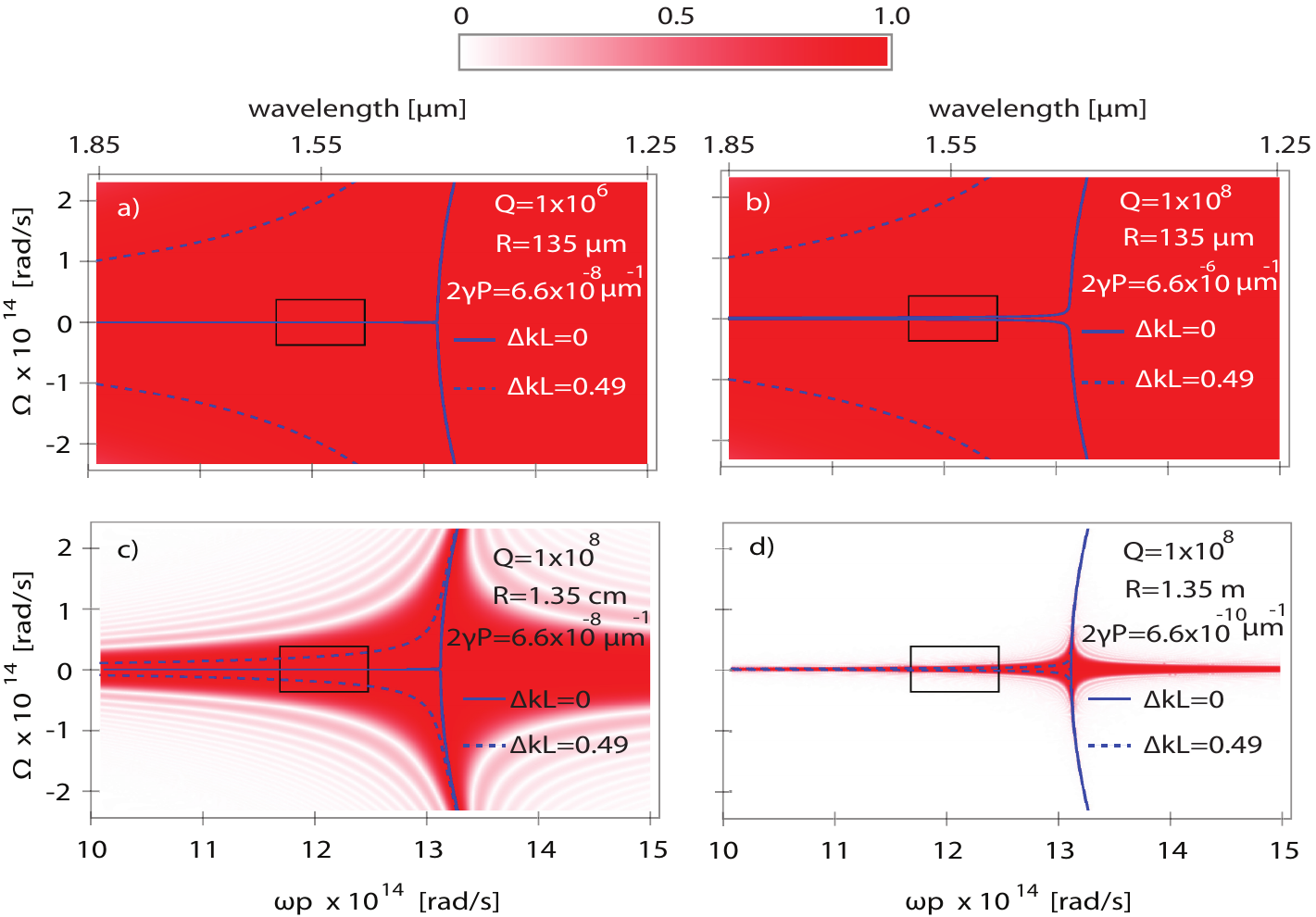}
	\caption{(a) and (b): Density plot of the phasematching function  $\vert \mbox{sinc} (L \Delta \kappa_{\vec{l}_p  \vec{l}_s)  \vec{l}_i}(\omega_{p},\Omega)/2)\vert^2$, also shown are the specific contours defined by $L \Delta k=0$ and $L \Delta k=0.49$ (which correspond to $sinc (L \Delta k/2)$ equal to $1$ and $0.99$, respectively),  for a sphere with $R=135\mu m$, for two values of $Q$, as indicated.
		The black rectangle indicates the region of interest in our experiments, centered around 1550nm. (c) and (d): Similar plots to (a) and (b), for $Q=10^8$ and much larger radii ($R=1.35$cm and $R=1.35$m).}
	\label{fig:phase-matching-vs-g}
\end{figure*}

With the help of the above theory, we can now describe the spectral and temporal properties of the two photon state produced by SFWM in a specific source design.  In this section we present simulations of the photon-pair properties of interest, assuming experimental parameters consistent with our experiments described in Sec. \ref{sec:experiment}.
Let us consider a fused silica sphere of radius $R=135\; \mu$m. 
For this radius, the degenerate SFWM frequency $\omega_s=\omega_i=\omega_p=2 \pi c/1550.92nm$ corresponds most closely with azimuthal index values $l_s=l_i=774$ (i.e. this value of $l$ comes closest to fulfilling the resonance condition  $k(\omega_{l})L=2\pi l$).

In Fig. \ref{fig:phase-matching-vs-g} a) and b) we plot for two values of the $Q$ parameter (namely $Q=10^6$ and $Q=10^8$), for a radius of 135$\mu$m, the phasematching strength $\vert g_{\vec{l}_p \vec{l}_s \vec{l}_i}(\omega_{p},\Omega)\vert^2$, as a function of the pump frequency $\omega_p$ in the horizontal axis and the signal-photon generation frequency (detuned from the pump) $\Omega$ in the vertical axis. Note that fixing the input pump power to $P_{in}=7$mW, the $Q$ value defines $2 \gamma P$ through \eqref{equ:gamma}.  In each of the plots we also show the phasematching contours $L \Delta k=0$ and $L \Delta k= 0.49$, the latter value chosen because it yields $sinc(L \Delta k/2)\approx 0.99$. In addition, a rectangle appearing close to the center of each plot represents the pump / generation  spectral region of interest in our experiments.  From these  plots we can draw the following two conclusions: i) the phasematching term $ g_{\vec{l}_p \vec{l}_s \vec{l}_i}(\omega_{p},\Omega)$ is essentially constant within the spectral area of interest, ii) changes in the nonlinear term $2 \gamma P$ (within the experimental range of interest) do not affect the resulting photon-pair properties.   What this means is that we are able to approximate $ g_{\vec{l}_p \vec{l}_s \vec{l}_i}(\omega_{p},\Omega) \approx 1$, with the implication that the two-photon state will be entirely determined by the cavity properties through the Airy functions $\mathscr{A}_p(\omega)$, $\mathscr{A}_s(\omega)$, and $\mathscr{A}_i(\omega)$, as in the second line of \eqref{JSIfinal}.

In this context, so as to illustrate the effect of using extended rather than micro-cavities,  in Fig.  \ref{fig:phase-matching-vs-g} c) and d) we show, for cavity radii of $1.35$cm and $1.35$m and a $Q$ value of $Q=10^8$ the phasematching strength $\vert g_{\vec{l}_p \vec{l}_s \vec{l}_i}(\omega_{p},\Omega)\vert^2$, as a function of $\omega_p$ and  $\Omega$.  As in panels a) and b) we show with a rectangle the spectral area of interest in our experiments.   Note that in contrast with microcavities which are the focus of this work, for extended cavities phasematching does become a relevant factor in defining the two photon state.

\begin{figure*}
	\centering
	\includegraphics[width=17cm]{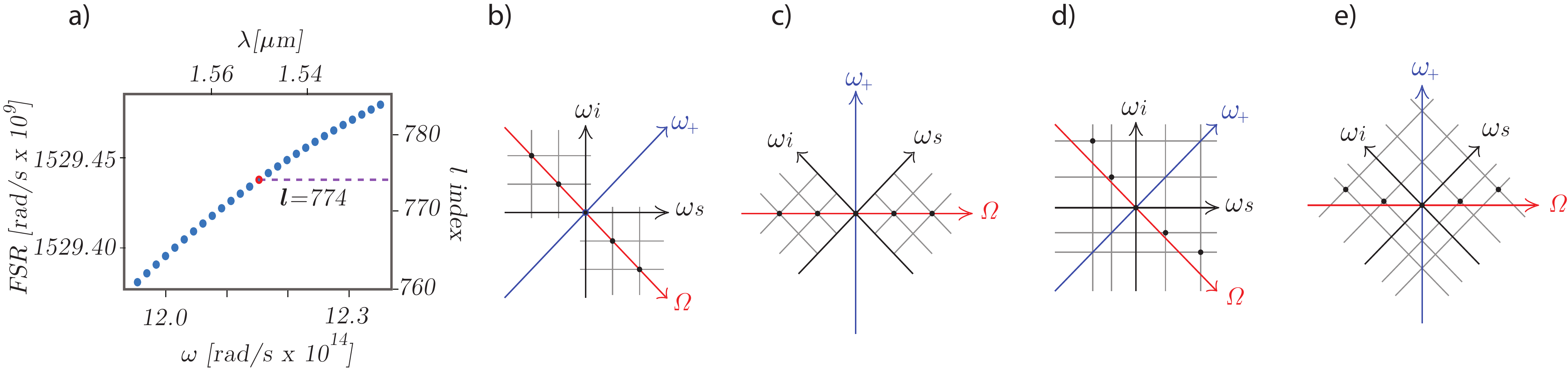}
	\caption{ a) FSR vs optical frequency dependence, calculated from \eqref{E:disp}.   b) and c)  Sketch of generation-mode matrix assuming a constant FSR, plotted vs $\omega_s$ and $\omega_i$ (in panel b)), and vs $\Omega$ and $\omega_+$ (in panel c)).     d) and e)  Sketch of generation-mode matrix assuming a FSR with  spectral drift, plotted vs $\omega_s$ and $\omega_i$ (in panel d)), and vs $\Omega$ and $\omega_+$ (in panel e)).}
	\label{fig:resonances-and-FSR}
\end{figure*}

Considering the discussion above, it is the cavity resonances for the signal and idler modes, as well as for the pump, which determine the two photon state. Note that along the signal frequency $\omega_s$  axis, the two-photon amplitude can be non-zero only within each of the resonances (each corresponding to a particular value of $l_s$), and likewise for the idler frequency $\omega_i$ axis.  Therefore, the two-photon state  can be non-zero in $\{\omega_s,\omega_i\}$ space only within each particular mode belonging to a matrix of modes, defined by a all combinations of $l_s$ and $l_i$, centered around $l_s=l_i=774$.

We point out that the spectral separation between two neighboring spectral modes, also known as free spectral range (FSR), has a slight dependence on frequency, as plotted in Fig. \ref{fig:resonances-and-FSR}  a) from \eqref{E:disp}.   The effect of such a spectral drift in the FSR on the two-photon state structure  is illustrated in the panels b)-e). In b) we present an illustration in $\{\omega_s,\omega_i\}$ space of  the matrix of generation modes under the assumption of a constant FSR.  We also indicate in this plot the rotated axes $\omega_{+}$ and $\Omega$, where a particular monochromatic pump corresponds to fixing the value of $\omega_{+}$ as $\omega_{+}=\omega_p$. An appropriate choice of $\omega_p$ so as to ensure overlap with the vertices of the resulting squares leads to a state in the form of a frequency comb along the main diagonal of the generation-mode matrix, as indicated in the figure.   In c) we show the same information in a rotated frequency space $\{\omega_{+},\Omega\}$, so that the generation modes appear along a horizontal line. Panels d) and e) are analogous to b) and c), except that we have included the effect of the spectral drift of the FSR. It can be appreciated that the two photon state structure now increasingly departs from the diagonal for large $\vert \Omega \vert$ (with a concave locus of all possible generation modes).

\begin{figure*}[h!]
	\centering
	\includegraphics[width=15cm]{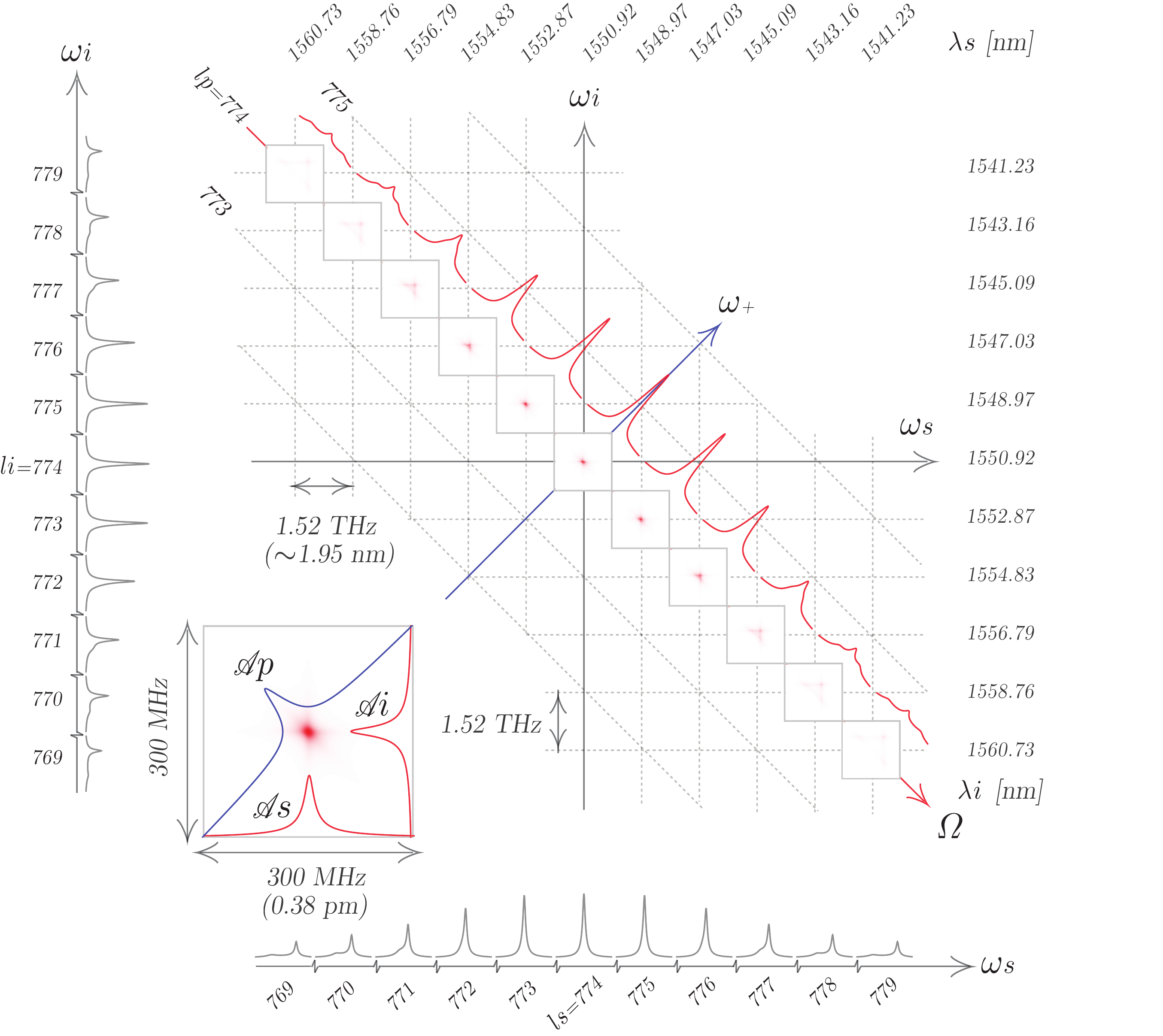}
	\caption{For a microsphere with radius $R=135\mu$m and $Q_s=Q_i=1\times10^{8}$ two-dimensional plot of the joint spectral intensity (computed from \eqref{Eq:2dJSI}), plotted within a region of 300MHz width, centered at each point of the generation mode-matrix main diagonal (with a separation of $1.52$THz between regions along $\omega_{s,i}$).  Inset: Close-up of central region, corresponding to $l_s=l_i=774$.}
	\label{fig:2dJSI}
\end{figure*}

In our two-photon state modelling, a prominent role is played by the resonance function for the pump $\mathscr{A}_p(\omega)$.   This function may be characterized experimentally, see section 4\ref{sec:4A}.  While in Fig. \ref{fig:resonances-and-FSR} we present sketches  designed to explain the geometry of the resulting SFWM two-photon state, in Fig. \ref{fig:2dJSI} we present an actual plot of the joint spectral intensity $\vert f_2(\omega_s,\omega_i;\omega_p) \vert^2$ (see \eqref{Eq:2dJSI}) for the specific case of a sphere of $R=135 \mu$m radius, assuming $Q$ values for the signal and idler modes of $Q_s=Q_i=1\times10^{8}$, and a full-width at half maximum width for function $\mathscr{A}_p(\omega)$ of 20.4MHz. Note that we have plotted the joint spectral intensity at each location on the main diagonal of the generation-modes matrix, within a square of $300$MHz side, and a center-to-center square separation of $1.52$THz. In this plot we have included along each of the $\omega_s$,  $\omega_i$, and  $\Omega$ (diagonal)  axes the resulting marginal intensity distributions.  In the inset, we show the central square corresponding to $l_s=l_i=774$, where we also include plots of $\mathscr{A}_s(\omega)$,  $\mathscr{A}_i(\omega)$, and $\mathscr{A}_p(\omega)$, along each of the $\omega_s$, $\omega_i$, and $\omega_+$ axes.

\begin{figure*}[h]
	\centering
	\includegraphics[width=15cm]{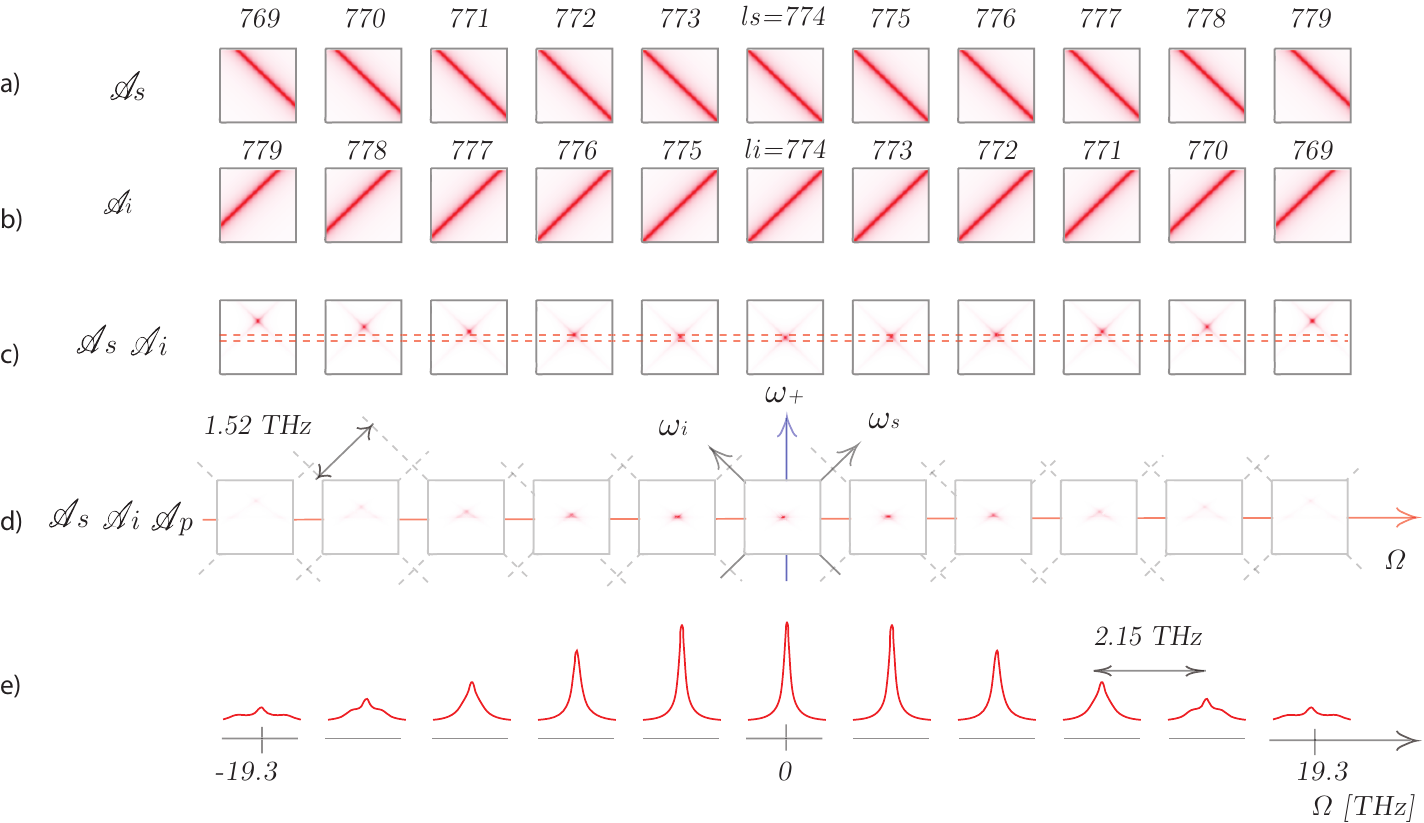}
	\caption{This figure clarifies the structure of the two-photon state produced by SFWM in a microsphere device, in regions similar to those defined in Fig. \ref{fig:2dJSI} for the same source, in the frequency variables $\Omega$ and $\omega_+$.  We show plots of the function $\mathscr{A}_s(\omega_
		++\Omega)$,  $\mathscr{A}_i(\omega_+-\Omega)$, $\mathscr{A}_s(\omega_
		++\Omega)\mathscr{A}_i(\omega_+-\Omega)$, and $\mathscr{A}_s(\omega_
		++\Omega)\mathscr{A}_i(\omega_+-\Omega) \mathscr{A}_p(\omega_+)$ in each of rows a) through d).  Note that in c) we have indicated with a pair of dotted lines the spectral width of the pump resonance.   In e) we show a plot of the function $R_i(\Omega)$, obtained by integrating $\mathscr{A}_s(\omega_
		++\Omega)\mathscr{A}_i(\omega_+-\Omega) \mathscr{A}_p(\omega_+)$ over $\omega_+$ (or equivalently over $\omega_p$; see \eqref{JSIfinal}), yielding the idler-photon spectral intensity in the form of a frequency comb. }
	\label{fig:2dJSIrot}
\end{figure*}

In order to further clarify the two-photon state structure, we plot in Fig. \ref{fig:2dJSIrot} within similar square regions as in Fig. \ref{fig:2dJSI} (this time defined in the rotated variables $\{\Omega,\omega_+\}$), the function $\mathscr{A}_s(\omega_++\Omega)$ in row a), the function $\mathscr{A}_i(\omega_+-\Omega)$ in row b), the product  $\mathscr{A}_s(\omega_++\Omega)\mathscr{A}_i(\omega_+-\Omega)$
in row c), and the product $\mathscr{A}_s(\omega_++\Omega)\mathscr{A}_i(\omega_+-\Omega) \mathscr{A}_p(\omega_+)$ in row d). Here, we may appreciate the effect already discussed through the sketches in Fig. \ref{fig:resonances-and-FSR}, in which the generation modes increasingly depart from the axis $\omega_+=\omega_p$ for larger values of $\vert \Omega \vert$.  In row c), we also indicate with two dotted lines the width of the pump resonance function $\mathscr{A}_p(\omega_+)$.   The result is that the generation modes lie increasingly outside of the pump resonance, and are therefore suppressed for large $\vert\Omega\vert$, as is clear in row d). In row e), we show the resulting photon-pair spectral intensity, in the form of a frequency comb, as is obtained from \eqref{JSIfinal}.

\begin{figure*}[h]
	\centering
	\includegraphics[width=14cm]{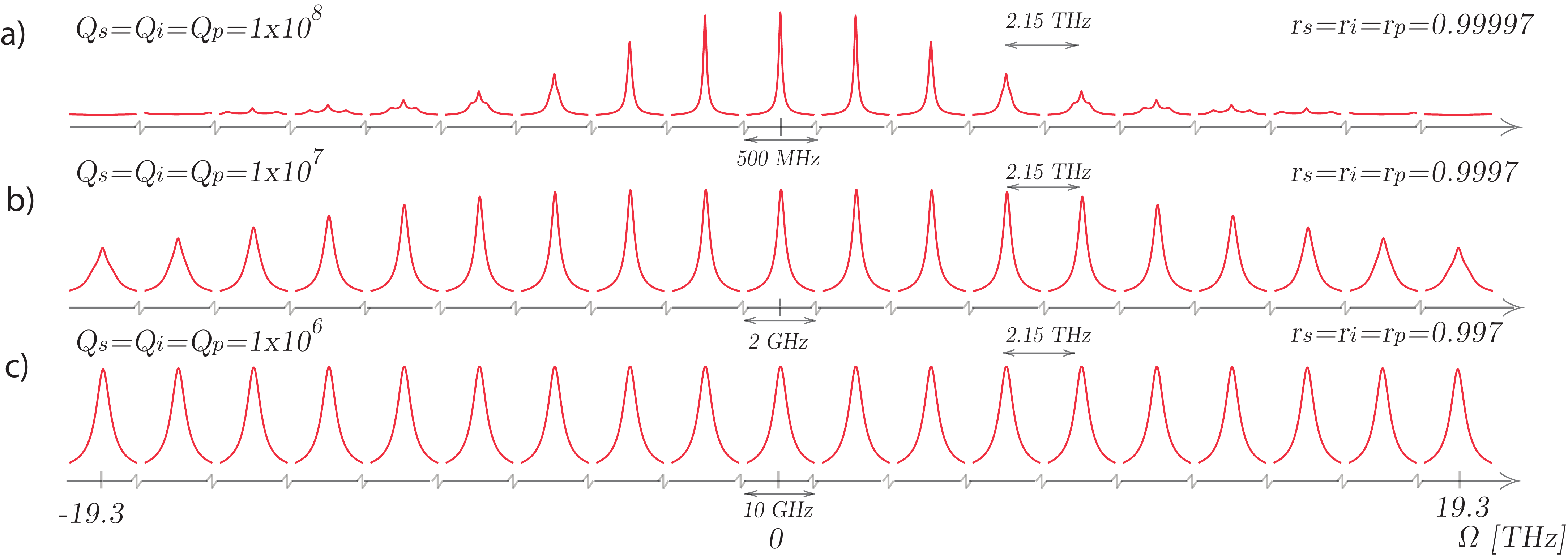}
	\caption{In this figure we show the behavior of the idler-photon spectral intensity $R_i(\Omega)$, for the same source as assumed in the previous two figures (computed from \eqref{JSIfinal}, see also Fig. \ref{fig:2dJSIrot}), for three different values of the $Q$ coefficients, assumed to be the same for both signal and idler modes.}
	\label{fig:combs}
\end{figure*}

In Fig. \ref{fig:combs} we show the idler-photon spectral intensity, similar to that shown in Fig. \ref{fig:2dJSIrot} e), for three different signal / idler $Q$ values ($1\times 10^6$,$1\times 10^7$, and $1\times 10^8$), where we also indicate the corresponding cavity reflectivity coefficients. Note that for a comparatively smaller $Q$ parameter,  resulting in spectrally broader generation modes, the envelope decays more slowly, because spectral overlap is retained with the pump resonance function $\mathscr{A}_p(\omega)$ over a larger span of $\Omega$ values.   

\begin{figure*}[h]
	\centering
	\includegraphics[width=16cm]{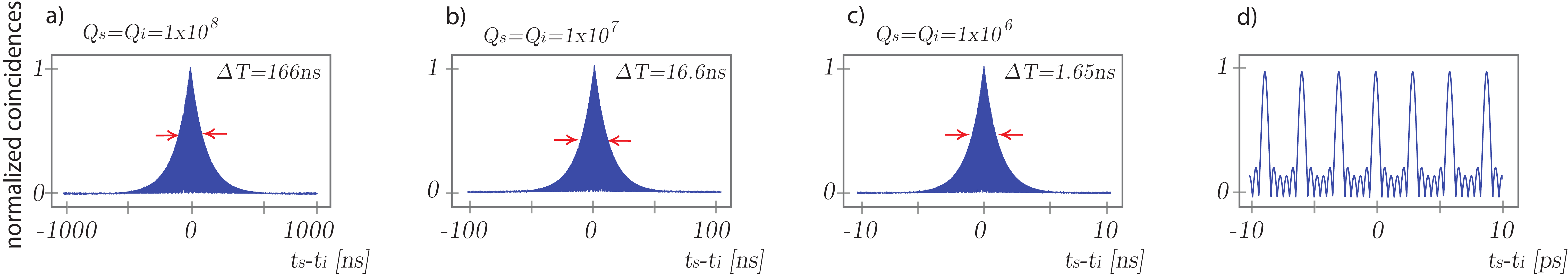}
	\caption{ In panels a) through c) of this figure we show plots of the temporal emission distribution (TED) function $\tilde{R}(T)$ (with $T=t_s-t_i$), obtained for the same source as assumed for the Figs. \ref{fig:combs} a)-c), for three different values of $Q=Q_s=Q_i$.   d) Closeup around the central region of the plot in a), showing the temporal oscillations with period equal to the cavity round trip time, which cannot be resolved in panels a) through c).}
	\label{fig:temporal_comb}
\end{figure*}

It is of interest to provide a temporal description of the two-photon state, in addition to the spectral description already provided.   In Fig. \ref{fig:temporal_comb} we show, as calculated numerically from \eqref{E:TED} and from the traces in Fig. \ref{fig:combs}, the time of emission distribution (TED) for the same three $Q$ values (panels a) through c)).   Note that each of these three functions exhibit an envelope with closely spaced oscillations (with a period corresponding to the cavity round trip time) which cannot be resolved in those plots.  In panel d), we show a closeup of the curve shown in panel a), within the region of the maximum showing these temporal oscillations.

As can be appreciated from Fig. \ref{fig:combs}, the resulting single-photon spectral intensity is in the form of a frequency comb, with the relative heights of the peaks modulated by an envelope function.  While the FSR exhibits a slow frequency dependence (as indeed has been shown in Fig. \ref{fig:combs} a)),  within a restricted spectral window we may model the SI function as a fixed-FSR comb function as follows

\begin{equation}
	R_i(\Omega)=H(\Omega) \cdot h(\Omega) \ast \mbox{comb}_{\delta \Omega} (\Omega).
	\label{equ:G}
\end{equation}

\noindent  where $H(\Omega)$ represents the envelope function, $h(\Omega)$ represents an individual peak in the comb, and the symbol $\ast$ denotes a convolution.  In addition, this equation is written in terms of the Dirac-delta comb function $\mbox{comb}_\Delta(x)$ defined as follows

\begin{equation}
	\mbox{comb}_\Delta(x)=\sum\limits_{j=-\infty}^{\infty} \delta(x- j \Delta).
\end{equation}

Let functions $\tilde{H}(T)$ and  $\tilde{h}(T)$ represent the Fourier transforms of $H(\omega)$ and $h(\omega)$, respectively.  Under the assumption that the function $\tilde{H}(T)$ is much narrower than function $\tilde{h}(T)$, we can then write the joint temporal intensity (or temporal emission distribution, TED, function) $\tilde{R}(T)$, with $T=t_s-t_i$, as follows

\begin{equation}
	\tilde{R}(T)=\tilde{h}(T) \cdot \tilde{H}(T) \ast \mbox{comb}_{1/\delta \Omega} (T)
	\label{equ:Rt}
\end{equation}

Thus, the joint temporal amplitude function is composed of a comb function in the temporal variable $T=t_s-t_i$ with individual comb peaks defined by the function $\tilde{H}(T)$, with a peak to peak separation of $1/\delta \Omega$, and an envelope function $\tilde{h}(T)$ which describes the roll-off in amplitude  for large $\vert T \vert$.   Interestingly, the roles of the function $H(\Omega)$ and $h(\Omega)$  in the frequency domain and $\tilde{H}(T)$ and  $\tilde{h}(T)$ in the temporal domain are reversed.     Specifically, the spectral envelope defines the functional dependence of each individual temporal comb peak, and the functional dependence of each individual frequency comb peak defines the envelope of the resulting comb in the temporal domain.   

Note also that if a spectral filter is applied in such a manner that a single spectral peak $R_i(\omega)=h(\omega)$ survives, we may obtain the functional dependence of $h(\omega)$ from a numerical Fourier transform of the envelope of the time of emission distribution $\tilde{h}(T)$.   This will be relevant in our experiment, below, in which while we lack the spectral resolution to resolve one comb peak $h(\omega)$ we can nevertheless  obtain this function $h(\omega)$ from a measurement in the temporal domain.

Note that it is possible to apply the converse of this idea:  from an experimental measurement of the spectral envelope $H(\Omega)$ we could infer through a Fourier transform the functional form of a single temporal peak $\tilde{h}(T)$.  We have not exploited this in our paper because the width of a single temporal peak is of less interest, as compared to the width of a single frequency comb peak, and because experimentally we do not obtain $H(\Omega)$ over the complete spectral range of interest, but only within the signal and idler spectral windows (see for example Fig. \ref{fig:wdm}(b) below).

\section{Experiment}
\label{sec:experiment}

In order to demonstrate the generation of sub-MHz spectral bandwidth photon pairs through the  spontaneous four wave mixing (SFMW) process, fused silica microspheres are used. The microspheres are fabricated from SMF-28 fiber using a fusion splicer (Fujikura FSM100P). The $135-250\mu$m radius spheres remain supported by an optical fiber stem, facilitating placement and alignment in our experimental setup.

The experimental setup used for our main results (Fig. \ref{fig:360b}, as well as Fig. \ref{fig:270a}) is  shown in Fig. \ref{fig:setup}.  Note that a number of setup variations, as indicated in the figures below, are used to obtain the various measurements reported, leading to our main results.

We have used as pump for the SFWM process a fiber-coupled, continuous wave laser, tunable within the wavelength range $1550-1630$nm, with a linewidth of $<200$kHz (New Focus TLB-6700). While the nominal linewidth is small, the laser can emit at a range of parasite frequencies. To remove extraneous modes, we have devised a filtering strategy involving a wavelength division multiplexing (WDM) device, followed by an erbium doped fiber amplifier, followed by a second WDM device. We use two WDM devices which we refer to as dense WDM (DWDM) on account of the comparatively small bandwidth of each channel, with a 0.57nm measured full width at half maximum (FWHM), as well as the comparatively small separation between channels (0.8nm), spanning wavelengths from 1529.55nm to 1560.61nm. With the amplifier operating at a gain of 24dB, the maximum usable laser power is 40mW; typically, the present experiments used less than 7mW.

\begin{figure*}[h!]
	\centering\includegraphics[width=14cm]{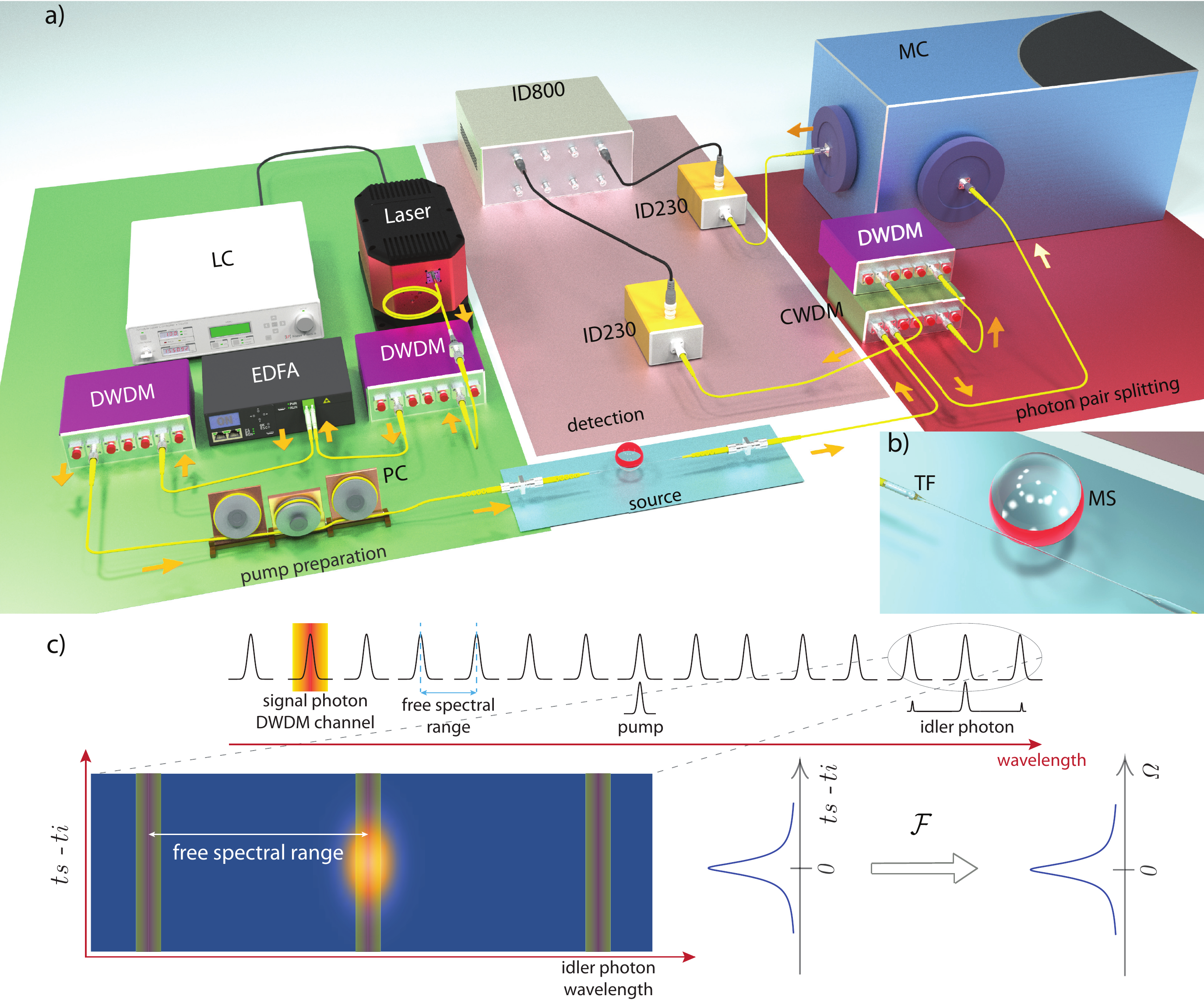} 
	\caption{\label{fig:setup} a) Experimental setup. LC: Laser controller, DWDM: `dense' wavelength division multiplexing device, EDFA: erbium-doped fiber amplifier, PC: polarization controller, TF:  tapered fiber, MS: microsphere, CWDM: 'coarse' wavelength division multiplexing device, MC: grating-based monochromator, ID230: free running InGaAs avalanche photodiode, ID800: time to digital converter.  Note that in some of the subsequent figures we have provided a setup sketch to show variations on the setup shown in this figure.  b) Close-up showing the fiber taper-microsphere system. c) Schematic of source operation and data obtained in our measurements.}
\end{figure*}

In order to couple the laser beam from the DWDM device into the microsphere, an evanescent fiber taper waveguide coupler is used. An inline optical polarization controller is used to adjust the polarization in the optical fiber before coupling into the cavity.  

\subsection{Cavity resonance characterization: setting up the SFWM pump}
\label{sec:4A}

\begin{figure}[h!]
	\centering\includegraphics[width=7cm]{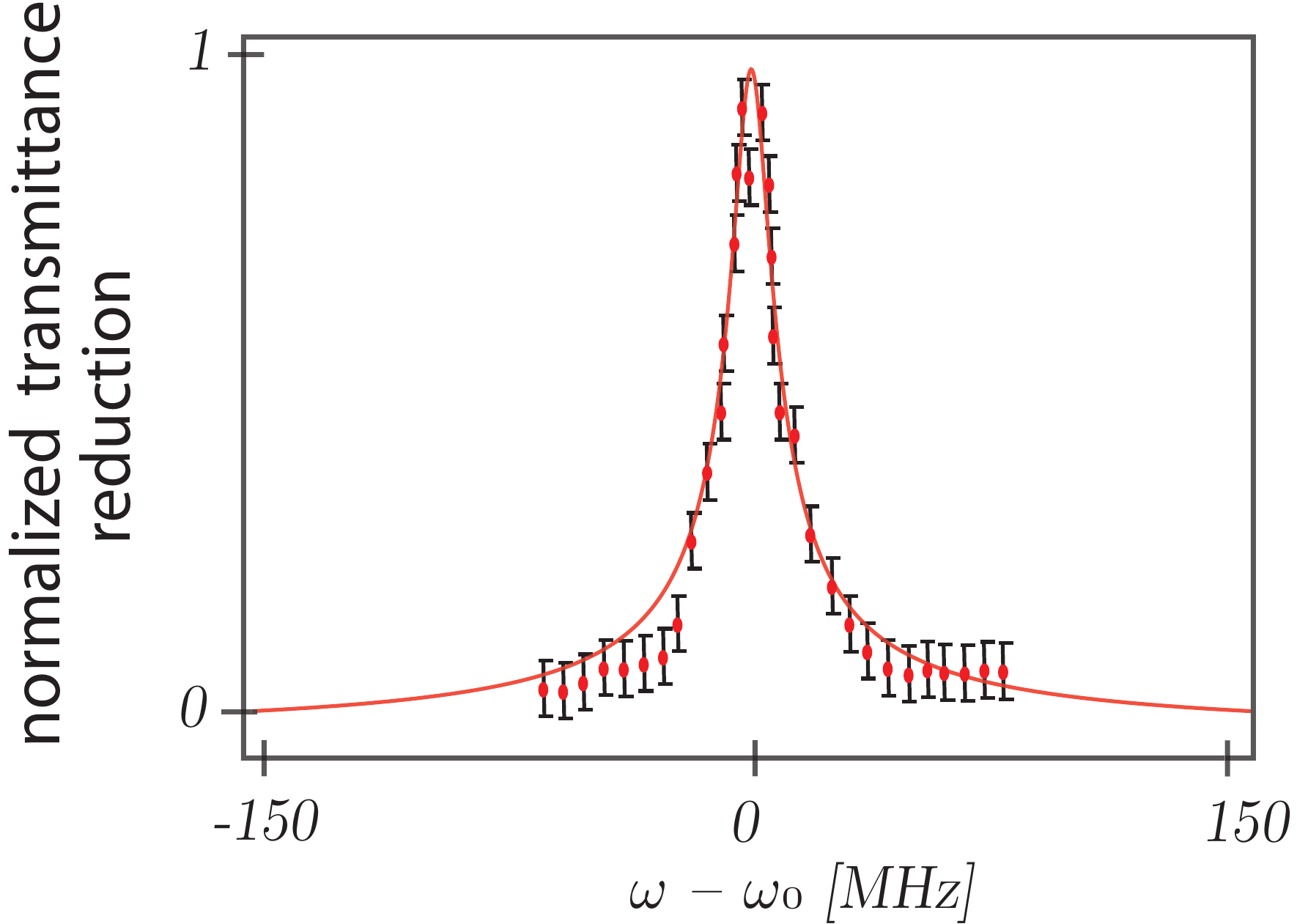}
	\caption{\label{fig:airy_ap} Experimentally-obtained normalized transmittance reduction vs $\omega$ (calculated as the difference between transmission far from the resonance and the transmittance at each $\omega$), for a particular cavity resonance, shown together with an Airy function $\mathscr{A}_p(\omega)$ fit. }
\end{figure}

In order to use our taper-sphere assembly, a first necessary step is to locate, and characterize, an appropriate resonance near 1550nm, which may act as pump for the SFWM process.  For this purpose, the laser is continuously rastered across a range of wavelengths, while monitoring the optical power emanating from the fiber taper output. The scan speed and time is optimized to eliminate any thermal effects which might distort the resonant lineshape.  In particular, we vary the laser frequency according to a triangular waveform with a 100Hz frequency and a 25GHz oscillation spread, while monitoring the transmitted power as recorded by a fast-photodiode (Thorlabs Model DET10D), with its electronic output leading to a 2GHz digital oscilloscope.
An example of a measurement obtained in this way is shown in Fig. \ref{fig:airy_ap}, where we plot the reduction in transmittance (transmittance at each frequency subtracted from the transmittance far from resonance) vs frequency. In addition to the experimental data, we also plot a best fit to an Airy  function $\mathscr{A}_p(\omega)$, which exhibits a full width at half maximum bandwidth of $20.4$MHz.   Note that multiple measurements over several resonances and a number of devices (with different radii) lead to a an approximate variation of  in the resonance bandwidth of $\sim 10\%$.

Once we have completed this step, we ensure that laser light can couple into the microsphere to act as pump in the SFWM process.  Note that thermal effects originating from the laser power confined in the microsphere result in a slight variation of the optical phase $ k(\omega)L$ over time, which is a sufficiently large effect to bring the microsphere out of resonance with the incoming laser frequency.  So as to circumvent this complication, we operate our experiments with the triangular variation of the laser frequency described above, thus ensuring that even if the system is brought out of resonance, it periodically returns to being on resonance at two points of each oscillating period.  Note that the oscillation spread (25GHz) is much greater than the width of the resonance (see Fig. \ref{fig:airy_ap}), which in turn is much greater than the pump linewidth.  We point out that our SFWM theory presented in section \ref{sec:theory} accounts for this pump frequency variation, by modelling the two-photon state as a statistical mixture of the pure states produced by each individual frequency within the pump resonance, as defined by function $\mathscr{A}_p(\omega)$.

\subsection{Single-photon frequency comb characterization}

While operating on-resonance, the microsphere presents a spectral comb of resonances. Note that the three waves involved in the SFWM process (degenerate pump, signal, and idler) must exhibit frequencies $\omega_p$, $\omega_s$, and $\omega_i$ matching one or more of these cavity resonances.  For a given combination of pump and signal-photon frequencies $\omega_p$ and $\omega_s$, the idler photon will appear at a frequency $\omega_i=2 \omega_p-\omega_s$, as mandated by energy conservation.

\begin{figure}[h!]
	\centering\includegraphics[width=7cm]{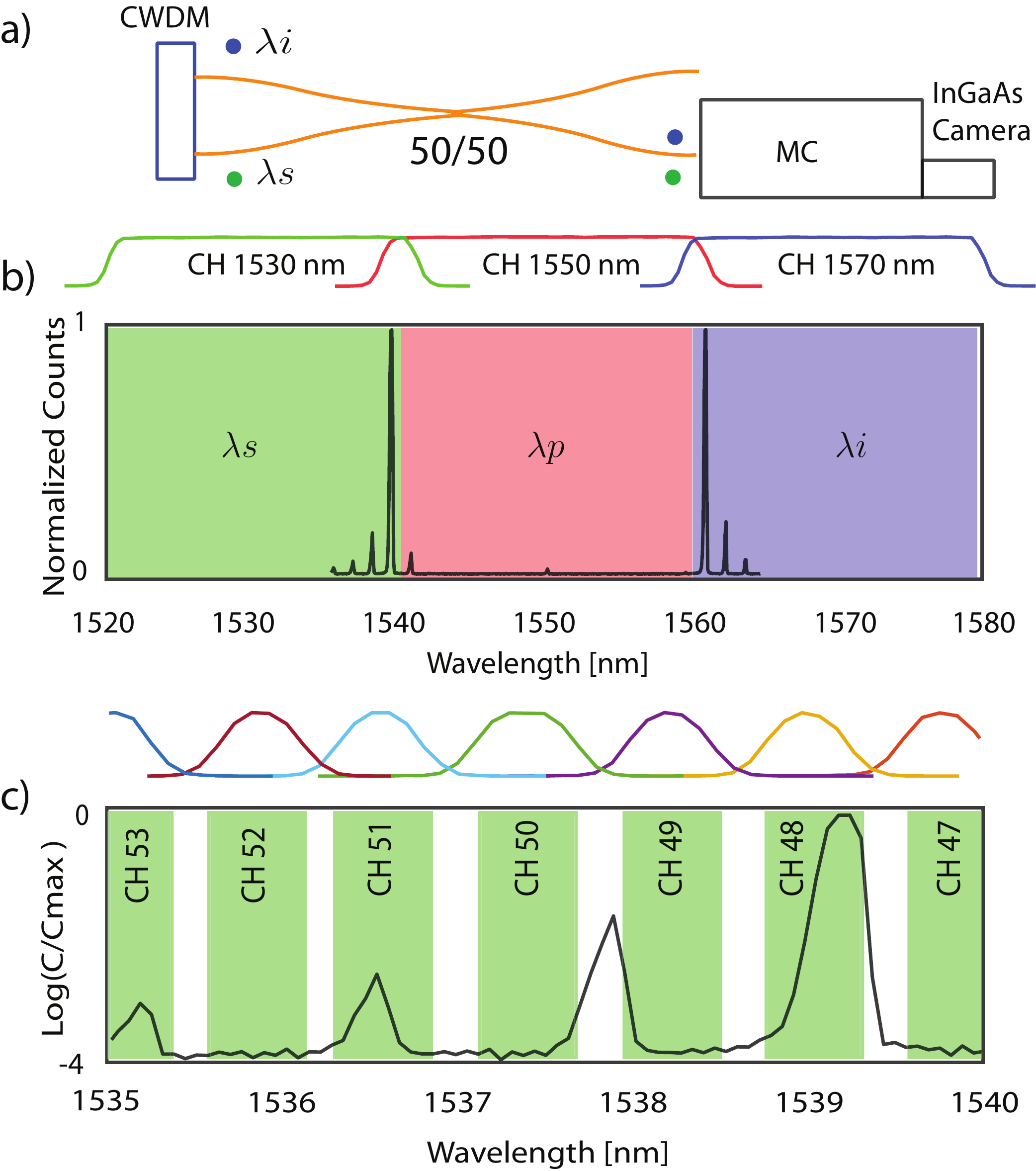}
	\caption{\label{fig:wdm} a) Sketch of the experimental setup used for the characterization of the signal-idler SFWM spectrum.  b) 
		SFWM spectrum, composed of pairs of energy-conserving peaks, also showing regions colored red, green, and blue indicating each of three relevant CWDM channels. We have included the measured transmission curves for these three CWDM channels.   c) For the signal photon, i.e. corresponding to $\lambda < \lambda_p$, the individual emission peaks (shown using a logarithmic scale), along with a map of the DWDM channels used, including the transmission curves for each of these channels.}
\end{figure}

In order to demonstrate the emission of SFWM photon pairs, we use the fact that the three waves involved (pump, signal, and idler) are all at different optical frequencies.  Thus, the output signal from the tapered fiber is sent to a coarse wavelength division multiplexing (CWDM) device which can send each of the three waves to a different output channel, as shown in the setup sketch in Fig. \ref{fig:wdm} a). This CWDM device has transmission windows with a FWHM measured width of 22nm, separated by 20nm (i.e. neighboring channels overlap), ranging from 1270nm to 1610nm.   Note that there are three outgoing CWDM channels which are of interest, centered at 1530nm, 1550nm, and 1570nm, as indicated with the colors  green, red  and blue, in Fig. \ref{fig:wdm} b), where we have also shown the experimentally-obtained channel transmission curves. Note that while the pump lies in the $1550$nm channel, the signal-photon band lies mostly in the 1530nm channel (red) and the idler-photon band lies mostly in the 1570nm channel (blue).   

We connect a single-mode fiber to each of the 1530nm and 1570nm channel outputs, thus splitting  the SFWM pairs into two distinct spatial modes, each travelling in a distinct fiber.  In order to visualize all SFWM generation peaks in a single spectral measurement, we first connect the two CWDM outputs into the two input ports of a 50:50 fiber beamsplitter, with one of the outputs leading to a grating spectrometer, with an InGaAs detection array (Andor Idus CCD camera) used as sensor; the experimental setup is skectched in Fig. \ref{fig:wdm} a).  The photon pairs are generated with the signal band roughly at 1540nm and the idler band roughly at 1560nm.    The result is the black curve in Fig. \ref{fig:wdm} b) showing three pairs of energy conserving peaks, along with two additional smaller peaks on the low-wavelength channel without an evident counterpart in the large-wavelength channel.  

\begin{figure*}[h!]
	\centering\includegraphics[width=17cm]{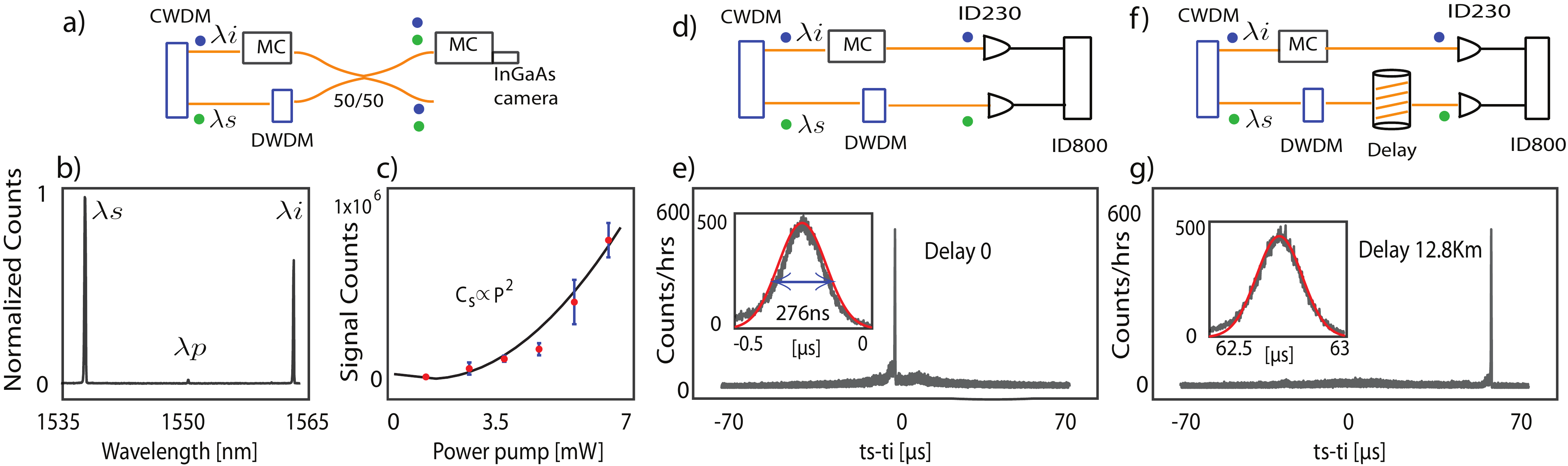}
	\caption{\label{fig:delays} a) Sketch of the setup used for the results shown in b) and c). b) Isolation of tallest pair of energy-conserving peaks from Fig. \ref{fig:wdm} a), obtained by filtering the signal photon with a DWDM channel and the idler photon with a grating monochromator. c) Total SFWM flux, integrated within the signal-mode peak in b), as a function of pump power, along with a quadratic fit. d) Sketch of the setup used for the results shown in e).  e) Coincidence count rate as a function of the detection time difference between the signal and idler modes, presenting  a well-defined peak near zero delay; inset: peak closeup.  f) Sketch of the setup used for the results shown in g). g) Similar to panel e), but the signal photon is transmitted through a 12.8km stretch of fiber.}
\end{figure*}

It becomes clear from the frequency comb obtained for each of the signal and idler photons, see Fig. \ref{fig:wdm} b), that there is one dominant pair of peaks (in terms of peak height). With the purpose of spectrally isolating this dominant pair of peaks,  we filter the signal photon  with a DWDM channel.  In Fig. \ref{fig:wdm} c) we show the same four peaks on the $\lambda<\lambda_p$ side which are transmitted by the $1530$nm CWDM channel (plotted in a logarithmic scale), together with the available DWDM spectral windows.  Thus, by transmitting the signal photon through DWDM channel 48, we can indeed isolate the tallest signal-photon peak.  In the next subsection, we perform an analysis of the coincidence count rate between the signal photon transmitted through the 1530nm CWDM channnel \emph{and} through DWDM channel 48, with the idler photon transmitted through the 1570nm CWDM channel.

\subsection{Initial coincidence count rate analysis}

The pairs of peaks described in the previous subsection are energy-conserving  which suggests that they are produced by a SFWM process. However, the photon-pair nature of the emitted light can be confirmed through a coincidence-counting measurement between the signal and idler photons.  For this purpose, as has already been mentioned, we isolate the tallest pair of peaks shown in Fig. \ref{fig:delays} a) by transmitting the signal photon  ($\lambda < \lambda_p$)  through  channel 48 of a DWDM (identical to the ones described above for filtering the pump). In addition, we transmit the idler photon ($\lambda > \lambda_p$)  through a grating monochromator, with its transmission window  centered at the energy-conserving idler-mode frequency $2\omega_p - \omega_s$.

In order to verify the energy conservation in this isolated pair of peaks, the DWDM-filtered signal photon (channel 48) is combined with the monochromator-filtered idler photon at a 50:50 fiber beamsplitter, and one of the outputs is sent to a second monochromator with its output leading to an InGaAs detection array (Andor Idus one dimensional CCD camera); the setup used is shown in Fig. \ref{fig:delays} a).  The result of this measurement is presented in Fig. \ref{fig:delays} b), clearly showing a single pair of energy-conserving peaks. At this point, we also performed a SFWM counts vs pump power measurement by numerically integrating the counts contained in the signal ($\lambda < \lambda_p$) photon as a function of the pump power. The results are shown in Fig. \ref{fig:delays} c) along with a quadratic best fit, indicating that the emitted power indeed has a quadratic dependence on the pump power as is expected for the SFWM process \cite{garay2012theory}. 

In order to experimentally observe coincidence events between the spectrally-filtered photons in each pair, we use the setup sketched in Fig. \ref{fig:delays} d), connecting the two fibers carrying the signal and idler generation modes to the entrance ports of two free-running InGaAs avalanche photodiodes (APD; IDQuantique 230). The electronic pulses produced by the APDs are sent to a time to digital converter (IDQuantique ID800) so as to monitor the distribution of signal-idler time detection differences. The results show a very well-defined peak near zero signal-idler delay (Fig.  \ref{fig:delays} e)). This peak corresponds to the detection of the idler photon, conditioned by the detection of the corresponding signal photon transmitted through the selected DWDM channel. This peak has a full width at half maximum of $276$ns and shows a temporal shift of $278$ns, which we believe is due to an imbalance in the $Q$ factor of the microsphere for the signal and idler photons. In other words, the cavity lifetime for the signal photon is longer than that of the idler photon by a time duration roughly corresponding to the width of the time of emission distribution. As an additional test, we performed a related experiment in which we transmit the signal photon over a 12.8km length of optical fiber, acting as a delay line, before reaching the corresponding APD Fig.  \ref{fig:delays} f). The resulting distribution of time of emission differences is presented in Fig.  \ref{fig:delays} g) and shows a temporal shift ($62.7\mu $s) corresponding to the delay due to the $12.8$km length of optical fiber.

As we have seen, DWDM channel 48 acting on the signal photon can 
transmit the tallest signal-frequency comb peak while suppressing all other peaks.   Thus, when monitoring \emph{coincidence} events as a function of the idler frequency (with the help of a grating-based monochromator), likewise only the tallest idler peak survives.  However,  the minimum frequency step $\delta \omega_{MC} \approx 39$GHz which can be resolved by the monochromator is orders of magnitude larger than the spectral width of the single photons (in the region of MHz), determined by the large $Q$ values of our resonator.  This means that while the monochromator is useful in order to spectrally situate the idler photon to within $\delta \omega_{MC}$, we are in fact unable to resolve the functional dependence of an individual frequency comb peak.  In the next subsection we will discuss a strategy based on temporally-resolved detection to overcome this limitation.

\subsection{Time-resolved and frequency-constrained coincidence count rate analysis}

As we have already discussed, in the context of Figs.  \ref{fig:delays} e) and g), we are able to measure the time of emission distribution envelope (corresponding to $\tilde{h}(T)$; see \eqref{equ:Rt}), with characteristic times in the hundreds of ns. 


As discussed at the end of section \ref{sec:example}, under the assumption that a single peak in the frequency comb can be isolated (although not resolved), we may recover the functional dependence of a single idler-photon comb peak, i.e. $h(\Omega)$, through the Fourier transform of the function $\tilde{h}(T)$ obtained experimentally.  In Fig. \ref{fig:setup} c) we present a schematic of the overall operation of our experiment: each of the pump, signal, and idler modes appear at one of the cavity resonances (i.e. are aligned with one of the cavity frequency comb peaks); the detection of a signal photon with frequency $\omega_s$, transmitted through a DWDM channel, heralds an idler photon, which is detected following passage through the monochromator set to transmit a frequency $\omega_i=2 \omega_p-\omega_s$ (thus, while sweeping $\omega_i$, a peak appears at the energy-conserving  value);  the one dimensional time of emission distribution (TED) envelope $\tilde{h}(T)$ obtained for a fixed idler transmission center frequency is Fourier transformed to yield the functional dependence of one idler-photon comb peak $h(\Omega)$.

\begin{figure*}[htp!]
	\centering
	\includegraphics[width=17cm]{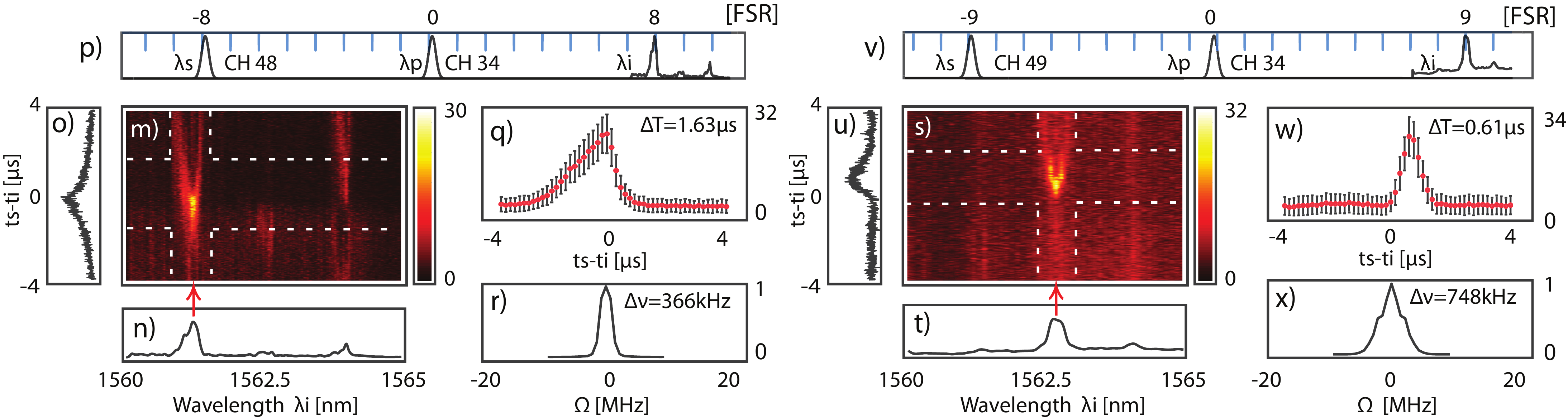}
	\includegraphics[width=17cm]{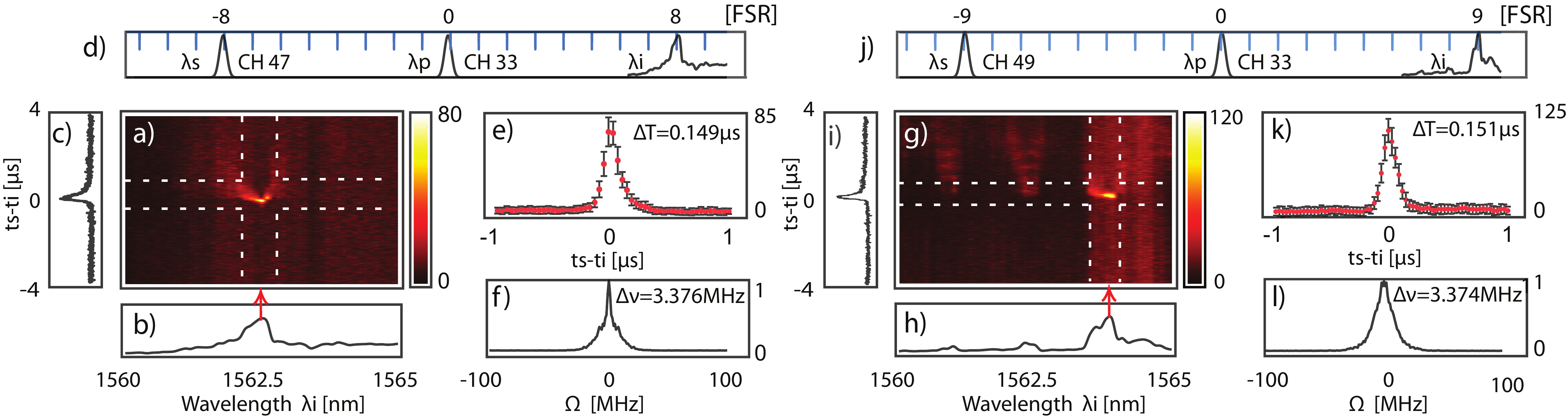}
	\caption{\label{fig:360b} For a microsphere with  $R=180\mu$m, the pump transmitted through DWDM channel 33, and the signal photon transmitted through DWDM channel 47: a) Heralded idler photon emission characteristics in the space formed by the idler wavelength $\lambda_i$ and the signal-idler time of detection difference $T=t_s-t_i$, b) marginal distribution in $\lambda_i$, c) marginal distribution in variable $T$, d) graphical representation of wavelengths involved for the pump, signal and idler modes, for idler frequency denoted by red arrow in a). e) Measured time of emission  distribution. f) Inferred idler-mode single photon intensity spectral distribution. Panels g)-l) are similar to panels a)-f), except that DWDM channel 49 is used, instead of 47, to transmit the signal photon. For a microsphere with red$R=180\mu$m, the pump is transmitted through DWDM channel 34, and the signal photon is transmitted through DWDM channel 48. m) Heralded idler photon emission characteristics in the space formed by the idler wavelength $\lambda_i$ and the signal-idler time of emission difference $T$, n) marginal distribution in $\lambda_i$, o) marginal distribution in variable $T$, p) graphical representation of wavelengths involved for the pump, signal and idler modes, for idler frequency denoted by red arrow in m), measured time of emission difference distribution shown in q), inferred idler-mode single photon intensity spectral distribution shown in r). Panels s)-x) are similar to panels m)-r), except that DWDM channel 49 is used, instead of 48, to transmit the signal photon.}
\end{figure*}

Following the strategy explained above, we carry out a measurement in which we spectrally filter the signal photon (with a DWDM channel) as well as the idler photon (i.e. we constrain it to the spectral transmission window of the monochromator) and, in addition, we temporally resolve each coincidence event, while sweeping $\omega_i$.   Specifically, we  measure the signal-conditioned idler time of emission  distribution as a function of the idler center frequency transmitted through the monochromator.  We thus obtain a two-dimensional density plot with the time of emission difference $T$ in the vertical axis and the idler wavelength $\lambda_i$ in the horizontal axis.  This type of measurement, carried out with the full setup shown in Fig. \ref{fig:setup}, leads to our main experimental results presented below.

The emission characteristics are dependent mainly on two experimental variables: the microsphere radius and the pump frequency. To investigate this dependence, we have conducted a systematic experimental study for distinct radius - pump wavelength combinations.

First, for a microsphere with $R=180 \mu$m, the pump transmitted through DWDM channel 33 ($\lambda=1550.92\pm0.25$nm), and the signal photon transmitted through channel 47 ($\lambda=1539.77\pm0.25$nm), Fig. \ref{fig:360b} a) shows the resulting two dimensional spectrogram. For the SFWM process, the observation of conditioned idler photons is expected at the energy-conserving frequency. In our measurement, we indeed observe a very well-defined peak centered at the energy conserving frequency, with residual accidental events appearing at multiples of the free spectral range (FSR = 1.4nm). Note that since we are not able to spectrally resolve the peaks, their apparent width is determined by the  effective spectral resolution in our setup, around $0.5$nm.  In Fig. \ref{fig:360b} b), we show a marginal spectral distribution obtained by numerically integrating over variable $T$ within the interval denoted by the dotted white lines, and in Fig. \ref{fig:360b} c), we show a marginal distribution in the time of emission difference $T$ obtained by numerically integrating over the frequency, again within the interval denoted by dotted white lines. The relevant pump and generation wavelengths are summarized in Fig. \ref{fig:360b} d), which shows the DWDM transmission windows used for the pump and signal modes, as well as the resulting measured spectral width of the idler photon. In this figure, we also indicate the microsphere resonance wavelengths.

It is of interest to measure the time of emission distribution (TED) envelope, corresponding to function $\tilde{h}(T)$ (see \eqref{equ:Rt}),  for a given filtering configuration, i.e., for a given choice of DWDM channel (signal photon) and a given  monochromator spectral  window (idler photon). Fig. \ref{fig:360b} e) shows the measured TED envelope for the signal photon transmitted through DWDM channel 47 and for the monochromator set to transmit the idler frequency $\omega_{i}=1206.37$THz, indicated with a red arrow in panel a). Fig. \ref{fig:360b} f) shows the single-photon spectral profile $ h(\Omega)$ of the heralded idler photon, obtained as explained in section \ref{sec:example}, from the Fourier transform of the TED, or $\tilde{H}(T)$. The observed bandwidth of the heralded idler photon (expressed in natural rather than angular frequency of $\Delta \nu=3.376$MHz is remarkably small, corresponding to a remarkably large idler photon $Q$ parameter (obtained as  $\omega_i /\Delta \Omega$ where $\omega_i$ is the central idler frequency) of $0.57 \times 10^8$.   Note that microspheres tend to exhibit a larger $Q$ parameter  as compared with other micro-resonator geometries (such as micro-rings\cite{reimer2015cross,grassani2016energy,kues2017chip,jaramillo2017persistent,preble2015chip,lu2019chip,silverstone2015qubit,caspani2016multifrequency} and micro-toroids\cite{rogers2019coherent}), which implies on the one hand a smaller attainable photon pair bandwidth, and on the other hand also implies that the SFWM photon pair emission rate for a given pump power level will  be higher.

In the SFWM process with a narrowband pump, we expect strict spectral correlations between the signal and idler photons, as indeed is implied by \eqref{E:2Pstate}. Therefore, shifting the signal photon filtering to a different wavelength (i.e. selecting a different DWDM channel), we expect a corresponding spectral shift for the idler photon as recorded by the monochromator-based measurement. The experimental verification of such spectral shifting due to signal-idler spectral correlations is presented in Fig. \ref{fig:360b} g)-l). Concretely, shifting the signal DWDM channel from 47 to 49 (corresponding to a central channel wavelength shift from $1539.77$nm to $1538.19$nm), results in a shift in the idler photon which is apparent in Fig. \ref{fig:360b} g), which is to be compared with Fig. \ref{fig:360b} a). In the remaining panels, we have shown analogous plots to those in the left-hand side of the figure: marginal spectral distribution for the idler photon in panel h), marginal time of emission difference distribution for the idler photon in panel i), summary of the spectral characteristics of the pump, signal, and idler modes in panel j), TED for a given filtering configuration in panel k), and single-photon spectral profile infered for the heralded idler photon in panel l). The resulting spectral width ($3.374$MHz) and $Q$ parameter ($0.56 \times 10^8$) values are similar to the ones in the earlier configuration.

The second part of Fig. \ref{fig:360b} contains a complementary set of data, for a different combination  pump wavelength. In the first combination, the pump wavelength is changed to $1550.12$nm, transmitted by DWDM channel 34 instead of 33, while leaving the microsphere radius the same. The qualitative behavior is similar to the earlier presented findings when the pump was in DWDM channel 33, and the heralded idler photon exhibits a considerably smaller emission bandwidth (again expressed in terms of natural rather than angular frequency) of 366kHz and 748kHz vs 3.376MHz and 3.374MHz, and larger $Q$ values, $5.3 \times 10^8$ and $2.5\times 10^8$ vs $0.57 \times 10^8$ and $0.56 \times  10^8$, as compared to the results with the pump at DWDM channel 33 (Fig. \ref{fig:360b}). We note that this represents, to the best of our knowledge, the shortest bandwidth for a heralded single photon (366kHz) demonstrated to date, based on the SFWM process.  

Note that in each two dimensional data set with axes $\{\lambda_i, T\}$, 10,000 experimental data points were recorded across the range of $T$ values; we grouped these data points in sets of either 100 (in the case of panels e) and k)) or 200 points (in the case of panels q) and w)), which were averaged to yield the one-dimensional TED functions with 100 data points for e and k and with 50 data points for q and w.  Error bars in each of the TED plots indicate the standard deviation among the 100 or 50 data points.    

As we have studied in sections \ref{sec:theory} and \ref{sec:example}, the resulting frequency comb envelope depends on the overlap of the functions $\mathscr{A}_p(\omega)$, $\mathscr{A}_s(\omega)$, $\mathscr{A}_s(\omega)$, which is affected by the spectral drift in the FSR discussed in section \ref{sec:example} (see Fig. \ref{fig:resonances-and-FSR} a).  We remark that the envelope width apparent in our experimental data shown in Fig. \ref{fig:360b} is broadly consistent with the simulated one in Fig. \ref{fig:temporal_comb}, for the experimental values of $Q$ obtained in our experiment.

Note that in addition to the smaller bandwidth, the emission characteristics in the $\{\lambda_i, T\}$ space include a pair of `tails' which extend towards positive $T$ values, the origin of which are left for a future study. In order to also show the idler-photon spectral shifting in response to shifting the signal-photon DWDM channel, we have filtered the signal photon with DWDM channel 48 in the left hand side of the figure and with channel 49 in the right-hand side. It is clear that the idler photon shifts in frequency in response to selecting a different signal-photon DWDM channel as we would expect for strict signal-idler spectral correlations.

\begin{figure}[h]
	\centering\includegraphics[width=8cm]{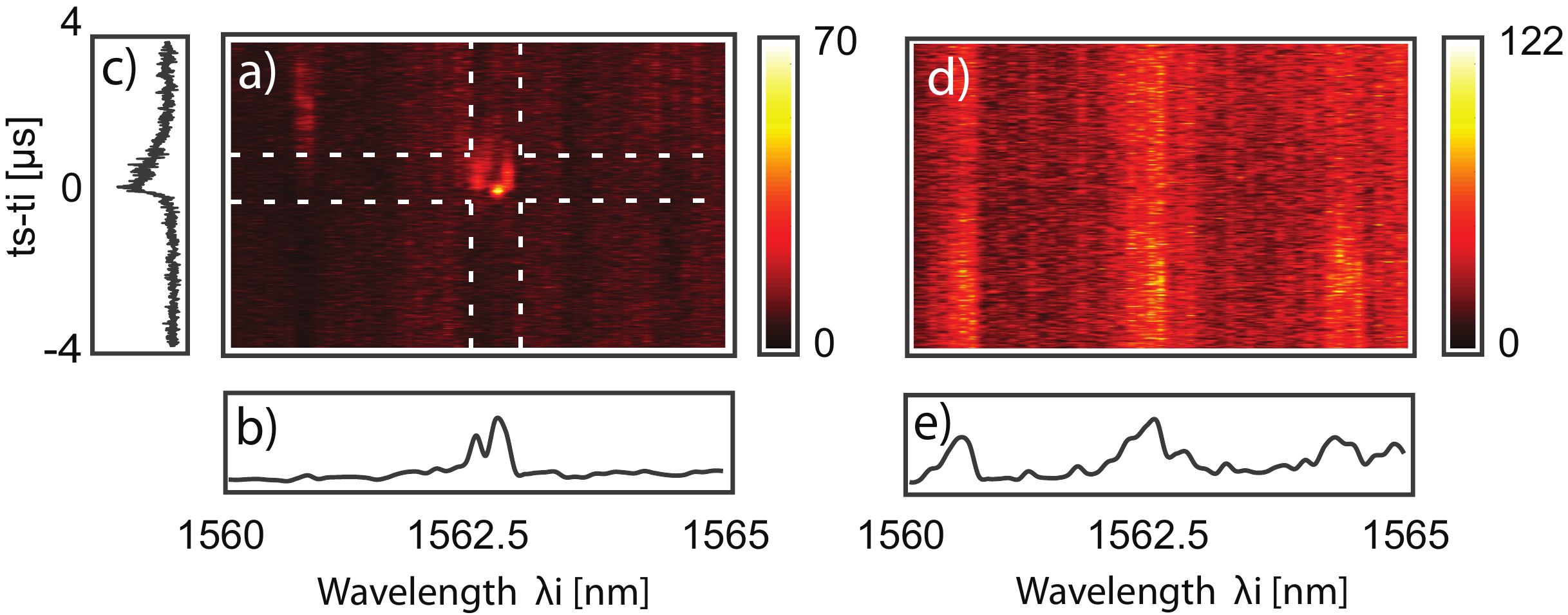}
	\caption{\label{fig:270a}  For a microsphere with $R=135\mu$m, and the pump transmitted through DWDM channel 33, a) and d) show the idler-mode, single-photon emission characteristics for narrowband filtering (DWDM channel 48) in a), and for wideband fitering (1530nm CWDM channel in d). Panels b) and e) show spectral marginal distributions for the distributions in a) and d), respectively.  Panel c)  shows a temporal marginal distribution for the distribution in a).}
\end{figure}

To investigate the dependence of the heralded idler photon on the type of signal photon filtering method used, two different filtering systems were applied: wide-band and narrowband. A slightly smaller radius microsphere was used red($135\mu$m), and the pump laser is adjusted to 1550.92nm, transmitted through DWDM channel 33. For this experiment we have used two signal-photon spectral filtering configurations: a narrowband configuration based on DWDM channel 48 (centered at $1538.98$nm with a $0.5$nm width), and a wideband configuration based on the CWDM channel only (centered at 1530nm with a 20nm width) see Fig. \ref{fig:270a}. Panels a) and d) show the heralded idler photon emission characteristics in the $\{\lambda_i, T\}$ space, under narrowband and wideband signal-photon filtering, respectively. For each of these panels, we have shown a marginal distribution for $\lambda_i$, obtained by integrating over $T$, see panels b) and e), as well as a marginal distribution for $T$, obtained by integrating over $\lambda_i$, see panel c). Note on the one hand that while the narrowband case results in a sharp peak, both in idler frequency and in time of emission difference, on the other hand the wideband case shows an essentially constant response along $T$ and three emission regions along $\lambda_i$ separated by the free spectral range, i.e. corresponding to thee micro-resonator spectral modes. It is consistent with the existence of signal-idler spectral correlations that a wider filtered spectral signal-photon span will map to a similarly wider idler-photon spectral generation range, as is observed experimentally. It is also noteworthy that the fact that the wideband case is not peaked in the time of emission difference variable $T$ suggests that the photon pair character is lost, probably as a result of noise mechanisms.

\begin{table*}[htbp]
	\begin{center}
		\begin{tabular}{|c c c c c c c|}
			\hline
			Radius [$\mu$m] & $\lambda_{p}$ [nm] & $\lambda_{s}$ [nm] & $\lambda_{i}^{(0)}$ [nm] & $\Delta T$ [$\mu$s]  & $\Delta \nu$ [MHz]  & $Q [\times10^{8}]$  \\
			\hline \hline
			i) 180  &   CH33 [1550.92]  & CH47 [1539.77] & 1562.30 & $0.149$  & \parbox{3em}{\centering 3.376}          & $0.56$   \\ \hline
			ii) 180  &  CH33 [1550.92]   & CH49 [1538.19] & 1563.70  & $0.151$ & \parbox{3em}{\centering 3.374}         & $0.57$   \\ \hline
			iii) 180  &   CH34 [1550.12]   & CH48 [1538.98] & 1561.31 & $1.63$  & \parbox{3em}{\centering 0.366}         &  $5.3$    \\ \hline
			iv) 180  &   CH34 [1550.12]  & CH49 [1538.19] & 1562.70 & $0.61$   & \parbox{3em}{\centering 0.748}          &  $2.5$    \\ \hline
			v) 135   &  CH33 [1550.92]   & CH48 [1538.98] & 1562.70 & $0.208$  &\parbox{3em}{\centering 2.370}           & $0.825$   \\ \hline
			vi) 135  &    CH33 [1550.92]   & CWDM [1530]      & not defined     & not defined        &  not defined   & not defined     \\ \hline
		\end{tabular}
		\caption{\label{tab:resumen} In this table we show details of the various experimental runs shown in Figs. \ref{fig:360b}-\ref{fig:270a}.   For each run we indicate te microsphere radius, the pump wavelength, the signal photon wavelength,  the value of the idler wavelength selected for displaying the TED, the width of the temporal distribution $\Delta T$ , the resulting bandwidth of the heralded idler photon, and the associated  $Q$ parameter. Rows i) through iv)  correspond to Fig. \ref{fig:360b}, while v) and vi) to Fig. \ref{fig:270a}. }
	\end{center}
\end{table*}

Table \ref{tab:resumen} shows a summary of all experimental runs discussed above.  For each specific experiment, we indicate the microsphere radius (in the first column), the central wavelength $\lambda_s$  of the DWDM channel used to filter the signal photon, or CWDM in the case of the last run (in the second column), the idler wavelength $\lambda_i^{(0)}$ selected in order to display the TED (in the fourth column), the width of the temporal distribution (in the fifth column), the resulting heralded idler photon spectral width $\Delta \nu$ (in the sixth column), and the associated $Q$ parameter (in the last column).  The first four rows correspond to the data shown in Fig. \ref{fig:360b}, while the fifth and sixth rows  to the data shown in Fig. \ref{fig:270a}.

We note that the taper-microsphere system has a considerable  complexity in terms of the supported modes.  The taper itself can support a number of transverse modes, and the pump light propagating in each of these can independently couple into a range of sphere modes, with an efficiency determined by an overlap integral (coupling coefficient) for each taper-sphere mode pair.   The pump circulating in each sphere mode then generates photon pairs propagating in turn in a range of sphere modes as determined by the intra-mode phasematching properties (see Eq.\ref{E:intramodalPM}).  These photon pairs then couple back to the taper, again in accordance with the specific coupling coefficients for each sphere-taper mode pair.    We point out that this description must then be applied to each pump frequency (rastered to eliminate thermal effects, see above), within the resonance bandwidth.

Note that while our theory includes the possible contribution of multiple sphere modes, and could be amended to  include multiple taper modes each coupling to a collection of sphere modes, in our numerical simulations  we assume for simplicity a single sphere mode participating in the SFWM process. Since we cannot at present control which sphere modes contribute to the detected two-photon state in each individual experimental run, an emission bandwidth with a considerable run-to-run variation is likely to result, as is in fact the case.

\section{Conclusions}

We have presented a photon-pair source based on the spontaneous four wave mixing (SFWM) process utilizing a fused silica microsphere as non-linear medium.  
In addition, we have presented a full theory for the SFWM process in these devices which fully takes into account all source characteristics relevant in our experiments. 
Our theory could be applied also to other types of cavities such as micro-rings and micro-disks, and predicts all important features of our experimental data including the $Q$-dependent shape and width of the single-photon frequency comb envelope. 
As a result of the high optical cavity $Q$ values obtained in our micro-spheres, heralded single photons with spectral widths down to $366$kHz are demonstrated. 
This represents a $\times 43$ improvement over previous work based on the SFWM process.  
We have measured SFWM spectra of the generated bi-photons, showing pairs of energy conserving peaks. 
Filtering a single pair of peaks, we have verified that the SFWM detection rate as a function of the pump power has a quadratic dependence as expected for the SFWM process and have shown a well-defined coincidence detection peak as a function of the signal-idler time of detection difference. 
We have presented a collection of coincidence count measurements as a function of the idler wavelength and the time of detection difference between the signal and idler modes, for a given choice of signal-mode spectral transmission window.  
These measurements represent the emission characteristics of an idler-mode heralded single photon, with the idler emission peaked at the expected energy-conserving wavelength and showing a wide time of detection difference distribution resulting from the cavity-enhanced SFWM process. 
While the spectral resolution in our measurements is useful to place the spectral peaks to to within $\sim 0.5$nm, we in fact unable to resolve the extremely narrow spectral widths of individual peaks in the single-photon frequency comb.
In this context we present an effective method for inferring the functional dependence of these individual peaks from an experimental measurement of the time of emission difference distribution.  
The ultra-narrow spectral linewidths made possible by the approach demonstrated here could enable a new set of applications for photon-based quantum information processing.

\section*{Funding}
This work was partially supported by CONACYT, Mexico (grants 293471,  293694,  Fronteras de la Ciencia 1667, 376135),  PAPIIT (UNAM) grant IN104418, AFOSR grant FA9550-16-1-1458 and Universidad de Guanajuato (CIIC Grant 028/2021).

\section*{Disclosures}

The authors declare no conflicts of interest.

\section*{Data Availability}

Data underlying the results presented in this paper are not publicly available at this time but may be obtained from the authors upon reasonable request.

\section*{Acknowledgments}
We acknowledge fruitful discussions with Andrea Armani, from the University of Southern California.

\printbibliography

\end{document}